\documentclass[longauth]{aa}

\usepackage{natbib}
\usepackage{graphicx}
\usepackage{txfonts}
\usepackage{array}
\usepackage{subfigure} 
\usepackage[skip=5pt]{caption} 
\usepackage[version=3]{mhchem}
\usepackage[export]{adjustbox}

\begin{document} 

   \title{Seeds of Life in Space (SOLIS). VI. Chemical evolution of sulfuretted species along the outflows driven by the low-mass protostellar binary NGC1333-IRAS4A} 
\titlerunning{Chemical evolution of sulfuretted species along the outflows driven by IRAS4A}

\author{
V. Taquet\inst{\ref{arcetri}}  
\and C. Codella\inst{\ref{arcetri},\ref{ipag}}  
\and M. De Simone\inst{\ref{ipag}} 
\and A. López-Sepulcre\inst{\ref{ipag},\ref{iram}}  
\and J. E. Pineda\inst{\ref{mpe}}  
\and D. Segura-Cox\inst{\ref{mpe}}  
\and C. Ceccarelli\inst{\ref{ipag},\ref{arcetri}}  
\and P. Caselli\inst{\ref{mpe}} 
\and A. Gusdorf\inst{\ref{ens},\ref{lerma}} 
\and M. V. Persson\inst{\ref{onsala}}  
\and F. Alves\inst{\ref{mpe}} 
\and E. Caux\inst{\ref{irap}} 
\and C. Favre\inst{\ref{ipag}} 
\and F. Fontani\inst{\ref{arcetri}} 
\and R. Neri\inst{,\ref{iram}} 
\and Y. Oya\inst{\ref{tokyo}} 
\and N. Sakai\inst{\ref{riken}} 
\and C. Vastel\inst{\ref{irap}} 
\and S. Yamamoto\inst{\ref{tokyo}}
\and R. Bachiller\inst{\ref{ign}}
\and N. Balucani\inst{\ref{perugia},\ref{arcetri},\ref{ipag}}
\and E. Bianchi\inst{\ref{ipag}}
\and L. Bizzocchi\inst{\ref{mpe}}
\and A. Chac\'on-Tanarro\inst{\ref{ign}}
\and F. Dulieu\inst{\ref{cergy}}
\and J. Enrique-Romero\inst{\ref{ipag}} 
\and S. Feng\inst{\ref{naoc}, \ref{naoc2}}
\and J. Holdship\inst{\ref{ucl}}
\and B. Lefloch\inst{\ref{ipag}}
\and A. Jaber Al-Edhari\inst{\ref{ipag},\ref{iraq}}
\and I. Jim\'enez-Serra\inst{\ref{jimenez}}
\and C. Kahane\inst{\ref{ipag}}
\and V. Lattanzi\inst{\ref{mpe}}
\and J. Ospina-Zamudio\inst{\ref{ipag}}
\and L. Podio\inst{\ref{arcetri}}
\and A. Punanova\inst{\ref{ural}}
\and A. Rimola\inst{\ref{rimola}}
\and I. R. Sims\inst{\ref{rennes}}
\and S. Spezzano\inst{\ref{mpe}}
\and L. Testi\inst{\ref{eso},\ref{arcetri}}
\and P. Theul\'e\inst{\ref{marseille}}
\and P. Ugliengo\inst{\ref{torino}}
\and A. I. Vasyunin\inst{\ref{ural},\ref{latvia}}
\and F. Vazart\inst{\ref{ipag}}
\and S. Viti\inst{\ref{ucl}}
\and A. Witzel\inst{\ref{ipag}} 
}

\institute{
INAF, Osservatorio Astrofisico di Arcetri, Largo E. Fermi 5, 50125 Firenze, Italy;
\email{taquet@arcetri.astro.it} \label{arcetri}
\and Univ. Grenoble Alpes, CNRS, IPAG, 38000 Grenoble, France \label{ipag}
\and Institut de Radioastronomie Millimétrique, 38406 Saint-Martin d’Hères, France \label{iram}
\and Max-Planck-Institut für Extraterrestrische Physik, Giessenbachstr. 1, 85748 Garching, Germany \label{mpe}
\and Laboratoire de Physique de l’ENS, ENS, Université PSL, CNRS, Sorbonne Université, Université de Paris, Paris, France \label{ens}
\and Observatoire de Paris, Université PSL, Sorbonne Université, LERMA, 75014 Paris, France \label{lerma}
\and Department of Space, Earth, and Environment, Chalmers University of Technology, Onsala Space Observatory, 439 92 Onsala, Sweden \label{onsala}
\and IRAP, Universit\'e de Toulouse, CNRS, CNES, UPS, Toulouse, France \label{irap}
\and Department of Physics, The University of Tokyo, 7-3-1, Hongo, Bunkyo-ku, Tokyo 113-0033, Japan \label{tokyo}
\and RIKEN Cluster for Pioneering Research, 2-1, Hirosawa, Wako-shi, Saitama 351-0198, Japan \label{riken}
\and IGN, Observatorio Astron\'omico Nacional, Calle Alfonso XII, 28004 Madrid, Spain \label{ign}
\and Dipartimento di Chimica, Biologia e Biotecnologie, Via Elce di Sotto 8, 06123 Perugia, Italy \label{perugia}
\and LERMA, Universit\'e de Cergy-Pontoise, Observatoire de Paris, 
PSL Research University, CNRS, Sorbonne Universit\'e, UPMC, Univ. Paris 06, 95000 Cergy Pontoise, France \label{cergy}
\and National  Astronomical  Observatory  of  China,  Datun  Road  20,Chaoyang, Beijing, 100012, P. R. China \label{naoc}
\and CAS  Key  Laboratory  of  FAST,  NAOC,  Chinese  Academy  of Sciences \label{naoc2}
\and Department of Physics and Astronomy, University College London, Gower Street, London, WC1E 6BT, UK \label{ucl}
\and University of AL-Muthanna, College of Science, Physics Department, AL-Muthanna, Iraq \label{iraq}
\and Centro  de  Astrobiología  (CSIC,  INTA),  Ctra.  de  Ajalvir,  km.  4,Torrejón de Ardoz, 28850 Madrid, Spain \label{jimenez}
\and Punanova's affiliation \label{punanova}
\and Departament de Qu\'{\i}mica, Universitat Aut\`onoma de Barcelona, 08193 Bellaterra, Catalonia, Spain \label{rimola}
\and Univ Rennes, CNRS, IPR (Institut de Physique de Rennes) - UMR 6251, F-35000 Rennes, France \label{rennes}
\and ESO, Karl Schwarzchild Srt. 2, 85478 Garching bei M\"unchen, Germany \label{eso}
\and Aix-Marseille Universit\'e, PIIM UMR-CNRS 7345, 13397 Marseille, France \label{marseille}
\and Universit\`a degli Studi di Torino, Dipartimento Chimica Via Pietro Giuria 7, 10125 Torino, Italy \label{torino}
\and Ural Federal University, 620002, 19 Mira street, Yekaterinburg, Russia \label{ural} 
\and Engineering Research Institute "Ventspils International Radio Astronomy Centre" of Ventspils University  \label{inst1}of Applied Sciences, Inženieru 101, Ventspils LV-3601,Latvia \label{latvia}
             }

   \date{}

 \abstract
 {
 Low-mass protostars drive powerful molecular outflows that can be observed with millimetre and submillimetre telescopes. Various sulfuretted species are known to be bright in shocks and could be used to infer the physical and chemical conditions throughout the observed outflows. 
 }
 {
 The evolution of sulfur chemistry is studied along the outflows driven by the NGC1333-IRAS4A protobinary system located in the Perseus cloud to constrain the physical and chemical processes at work in shocks.
 }
 {
 We observed various transitions from OCS, CS, SO, and SO$_2$ towards NGC1333-IRAS4A in the 1.3, 2, and 3mm bands using the {IRAM NOrthern Extended Millimeter Array (NOEMA)} and we interpreted the observations through the use of the Paris-Durham shock model. 
 }
 {
 The targeted species clearly show different spatial emission along the two outflows driven by IRAS4A. OCS is brighter on small and large scales along the south outflow driven by IRAS4A1, whereas SO$_2$ is detected rather along the outflow driven by IRAS4A2 that is extended along the north east - south west (NE-SW) direction. SO is detected at extremely high radial velocity up to $+25$ km s$^{-1}$ relative to the source velocity, clearly allowing us to distinguish the two outflows on small scales. 
 Column density ratio maps estimated from a rotational diagram analysis allowed us to confirm a clear gradient of the OCS/SO$_2$ column density ratio between the IRAS4A1 and IRAS4A2 outflows. 
 Analysis assuming non Local Thermodynamic Equilibrium of four SO$_2$ transitions towards several SiO emission peaks suggests that the observed gas should be associated with densities higher than $10^5$ cm$^{-3}$ and relatively warm ($T > 100$ K) temperatures in most cases.  
}
{
 The observed chemical differentiation between the two outflows of the IRAS4A system could be explained by a different chemical history. The outflow driven by IRAS4A1 is likely younger and more enriched in species initially formed in interstellar ices, such as OCS, and recently sputtered into the shock gas. In contrast, the longer and likely older outflow triggered by IRAS4A2 is more enriched in species that have a gas phase origin, such as SO$_2$.
 } 

   \maketitle
%

\section{Introduction}

The early stages of low-mass star formation are accompanied by the detection of bright bipolar low-velocity (typically $\sim$ 10 km s$^{-1}$) outflows on scales as large as several thousands of astronomical units (au).
These outflows can be produced by fast ($\sim$ 100 km s$^{-1}$) protostellar jets, which in turn form shocks {upon impact against} the high-density protostellar gas. {In addition, slower disc winds can also contribute in sweating up material. More specifically, \citet{Tabone2017} recently showed that SO and SO$_2$ emission as observed towards the HH212 protostar traces outflowing material launched by a region extended as the radius of the protostellar disc.}

Shocks are favourable regions to study how the chemical content of star forming regions is enriched thanks to sputtering (gas-grain collisions) and shattering (grain-grain) processes \citep{Draine1979, Tielens1994, Flower1994, Guillet2007}. In practice, the components of both dust mantles and refractory cores are injected into the gas. In addition, the sudden increases of temperature and density trigger a hot gas-phase chemistry otherwise not efficient in more quiescent regions.
As a consequence, the abundance of several molecular species dramatically increase by several orders of magnitude \citep{Bachiller2001, Herbst2009}. 

Ground-based or space infrared-(sub-)millimetre single-dish telescopes have been used for a few decades to probe the large-scale formation and evolution  of molecular outflows through the emission of bright species, such as CO, H$_2$O, NH$_3$, or CH$_3$OH, the main species of interstellar ices 
\citep[e.g.][]{Bachiller1995, Bachiller2001, Codella1999, Codella2005, Codella2010, Nisini2010, Kristensen2012, Yildiz2012, Busquet2014, Lefloch2017}.
On the other hand, interferometry is the way to image the small (down to $\sim$ 1$\arcsec$) regions chemically enriched due to the passage of a shock. In this context, a large number of studies have revealed bow shocks and cavities opened by protostellar jets \citep{Gueth1998, Bachiller1998, Benedettini2007, Benedettini2013, Codella2009, Codella2017, Persson2012, Santangelo2015, Lefevre2017}. 

\subsection{S-bearing species in outflows}

Many species showing a strong increase of their abundance in shocks along low-mass protostellar outflows are sulfuretted.
It has been assumed that H$_2$S is the main reservoir of sulfur in the dust mantles, its release in the gas phase through sputtering or thermal evaporation would then trigger the formation of gaseous SO followed by SO$_2$ \citep{PineaudesForets1993, Charnley1997}.
In other words, the H$_2$S/SO/SO$_2$ abundance ratios have been proposed as chemical clocks of the shocked gas. Unfortunately, the comparison with observations of S-bearing species along outflows did not provide any clear result in this context \citep{Codella1999, Wakelam2004}. In addition, so far H$_2$S has not been detected in interstellar ices \citep[e.g.][and references therein]{Boogert2015} questioning this hypothesis and calling for alternative possibilities. 
\citet{Wakelam2004} proposed that the sulfur released from dust mantles is in atomic form. On the other hand, OCS could be the solution. Although also OCS has been only tentatively detected on ices \citep{Boogert2015}, astrochemical models interpreting molecular surveys towards protostellar shocks along the { outflow driven by the L1157-mm Class 0 protostar (hereinafter L1157)} suggest that OCS represents at least 50\% of the sulfur reservoir in ice mantles \citep{Podio2014, Holdship2016}. 
All these findings call for more comprehensive datasets to shed light on the sulfuretted grain/ice composition.
A first step has been done by \citet{Holdship2019}, who analysed the unbiased spectral survey towards again L1157 performed with the IRAM-30m in the context of the Astrochemical Surveys At IRAM (ASAI) Large Program \citep{Lefloch2018}: more than 100 lines have been associated with the following molecules CCS, H$_2$CS, OCS, SO, SO$_2$, and H$_2$S \citep[from][]{Holdship2016} and collecting 10\% of the cosmic sulfur abundance.
It is time to perform a multiline analysis of sulfuretted species using interferometric data (to minimise filling factor effects) and towards other outflows than L1157 in order to check that the L1157, { a chemicallly rich outflow}, does not represent a unique case.

\subsection{The NGC1333-IRAS4A outflows}

NGC1333-IRAS4A (hereinafter IRAS4A) is a protobinary system located in the NGC1333 star-forming region of the Perseus cloud \citep[299$\pm$15 pc;][]{Zucker2018} { with a systemic velocity of 7 km s$^{-1}$} \citep{Yildiz2012}. The two Class 0 protostars are separated by 1$\farcs$8 \citep[538 au; see e.g.][]{Looney2000, LopezSepulcre2017, Tobin2018} and surrounded by an 8 $M_{\odot}$ envelope \citep{Maury2019}.
IRAS4A is one of the brightest { known} low-mass protostellar systems, with an overall internal luminosity of $\sim$ 5 $L_{\odot}$ \citep{Maury2019}, measured in the context of the {\it Herschel} Gould Belt survey \citep[see e.g.][]{Andre2010, Ladjelate2020}.

The two protostars have different characteristics: the SE source, called 4A1, shows the brightest mm-continuum emission, but it does not show any signature of chemically enriched protostellar cocoon whilst the NW protostar, 4A2, is associated with a hot-corino detected through various complex organic molecules \citep{Bottinelli2004, Bottinelli2007, Taquet2015, Santangelo2015, LopezSepulcre2017}.
The binary protostellar system drives two bipolar outflows observed on large scales (few arcminutes). \citet{Choi2005} mapped the outflows in SiO emission, mainly induced by the sputtering of refractory grains \citep[e.g.][and references therein]{Caselli1997, Gusdorf2008a, Gusdorf2008b}, at 2$\arcsec$ spatial resolution. Two different blue-shifted lobes are clearly observed towards South: more specifically, one along South and one towards south west. The northern lobe, as traced by Choi et al. (2005), shows only one lobe, which is abruptly bending toward north-west once at 20$\arcsec$ (6000 au) from the protostars.
Using IRAM-PdBI observations of CO, SO, and SiO lines with a spatial resolution of 300 au, \citet{Santangelo2015} were able to reveal two protostellar jets: a fast jet associated with H$_2$ shocked emission and driven by 4A1 (south lobe) and a slower and precessing jet emerging from 4A2 (SW). 
The northern red-shifted counterparts are not clearly disentangled even at interferometric scale.
Interestingly, interferometric CO images obtained with the SMA by \citet{Ching2016} suggest that the
magnetic field is wrapping the two northern red-shifted components. However, the complex red-shifted emission is far to be well revealed and imaged. This is crucial because the jet/outflow properties can be used to date the different evolutionary stages of the two protostars.
In practice, due to its fainter continuum emission, and its richer chemistry, 4A2 is thought to be older than 4A1. This picture is consistent with the { more elongated} outflow driven by 4A2 along the NE-SW direction relative to the outflow driven by 4A1 towards south.  

Here, we aim at analysing the emission of various transitions from four sulfuretted molecules, namely OCS, SO, SO$_2$, and CS with different excitation ranges and on different spatial scales along the outflows of IRAS4A.
An atomic sulfur abundance of $1-4 \times 10^{-7}$ towards the south outflow of IRAS4A has been estimated by  \citet{Anderson2013} from S[I] observations at 25 $\mu$m with {\it Spitzer}. 
We use interferometric observations taken in the context of the SOLIS \citep[Seeds Of Life In Space]{Ceccarelli2017} IRAM NOEMA Large Program, supported by further interferometric datasets.
The goal of this Large Program is to study the organic (and sulfuretted) molecular content in a sample of low-mass star-forming regions in different stages and environments. The spatial evolution of organic molecules along the outflows driven by IRAS4A is presented in a parallel and complementary study by \citet{DeSimone2019}.
The goal of the paper is twofold: 
%
%
Firstly, to derive for the first time 
the physical conditions of the IRAS4A outflows, so far observed only using
individual transitions of CO, SO, and SiO, and to disentangle the red-shifted 
northern components; 
%
Secondly, to investigate the evolution of the 
chemical conditions, using also up-to-date physical/chemical models of shocked regions, 
along the two outflows driven by IRAS4A. 


\section{Observations}

We make use of various Plateau de Bure Interferometer (PdBI) and NOrthern Extended Millimeter Array (NOEMA) datasets from four observational programs in the 1.3, 2, and 3mm bands and at different configurations. 
The properties of the transitions targeted in this work are summarised in Table \ref{transitions}. 

\begin{table*}[htp]
\centering
\caption{Properties of the transitions presented in this work.} 
\begin{tabular}{l c c c c c c c c}
\hline              
\hline              
Molecule	&	Transition	&	Frequency	&	$E_{\rm up}$	&	$A_{\rm ul}$	&	$\theta_{\rm beam}$	&	Dataset	&	Rms noise (blue)	&	Rms noise (red)	\\
	&		&	(GHz)	&	(K)	&	(s$^{-1}$)	&	($\arcsec \times \arcsec$)	&		&	(Jy beam$^{-1}$)	&	(Jy beam$^{-1}$)	\\
\hline																	
OCS	&	$8-7$	&	97.301209	&	21.0	&	2.58(-6)	&	$2.2 \times 2.0$	&	SOLIS-S3	&	7.40(-3)	&	5.57(-3)	\\
OC$^{34}$S	&	$7-6$	&	83.057970	&	15.9	&	1.59(-6)	&	$4.3 \times 3.4$	&	SOLIS-S1	&	5.70(-3)	&	5.80(-3)	\\
OCS	&	$17-16$	&	206.745155	&	89.3	&	2.55(-5)	&	$0.9 \times 0.7$	&	SOLIS-S5	&	1.53(-2)	&	1.18(-2)	\\
OCS	&	$18-17$	&	218.903355	&	99.8	&	3.03(-5)	&	$0.8 \times 0.7$	&	CALYPSO	&	2.00(-2)	&	1.73(-2)	\\
\hline																	
CS	&	$2-1$	&	97.980953	&	7.1	&	1.67(-5)	&	$2.2 \times 2.0$	&	SOLIS-S3	&	2.00(-1)	&	1.00(-1)	\\
C$^{34}$S	&	$2-1$	&	96.41295	&	6.2	&	1.60(-5)	&	$2.2 \times 2.0$	&	SOLIS-S3	&	1.06(-2)	&	5.50(-3)	\\
C$^{34}$S	&	$3-2$	&	144.617109	&	11.8	&	5.78(-5)	&	$2.1 \times 1.7$	&	T15	&	3.76(-2)	&	3.68(-2)	\\
\hline																	
SO	&	$3_2 - 2_1$	&	99.299870	&	9.2	&	1.12(-5)	&	$2.2 \times 2.0$	&	SOLIS-S3	&	7.90(-2)	&	4.20(-2)	\\
$^{34}$SO	&	$3_2 - 2_1$	&	97.715317	&	9.1	&	1.07(-5)	&	$2.2 \times 2.0$	&	SOLIS-S3	&	5.10(-3)	&	4.30(-3)	\\
SO	&	$6_5-5_4$	&	219.949442	&	35.0	&	1.36(-4)	&	$0.8 \times 0.7$	&	CALYPSO	&	9.50(-2)	&	8.40(-2)	\\
SO	&	$4_5-3_4$	&	206.176005	&	38.6	&	1.00(-4)	&	$0.9 \times 0.7$	&	SOLIS-S5	&	2.40(-2)	&	2.30(-2)	\\
\hline																	
SO$_2$	&	$5_{2,4}-5_{1,5}$	&	165.144651	&	23.6	&	3.12(-5)	&	$2.4 \times 1.8$	&	T15	&	4.90(-2)	&	5.30(-2)	\\
SO$_2$	&	$7_{1,7}-6_{0,6}$	&	165.225452	&	27.1	&	4.13(-5)	&	$2.4 \times 1.8$	&	T15	&	7.30(-2)	&	9.60(-2)	\\
SO$_2$	&	$8_{1,7}-8_{0,8}$	&	83.688093	&	36.7	&	6.83(-6)	&	$4.3 \times 3.4$	&	SOLIS-S1	&	1.20(-2)	&	1.05(-2)	\\
SO$_2$	&	$11_{2,10}-11_{1,11}$	&	205.300570	&	70.2	&	5.31(-5)	&	$0.9 \times 0.7$	&	SOLIS-S5	&	8.70(-3)	&	5.30(-2)	\\
SO$_2$	&	$13_{2,12}-13_{1,13}$	&	225.153703	&	93.0	&	6.52(-5)	&	$1.1 \times 0.8$	&	P12	&	4.02(-2)	&	3.29(-2)	\\
SO$_2$	&	$14_{1,13}-14_{0,14}$	&	163.605530	&	101.8	&	3.01(-5)	&	$2.4 \times 1.8$	&	T15	&	3.40(-2)	&	5.50(-2)	\\
SO$_2$	&	$16_{2,14}-16_{1,15}$	&	143.057110	&	137.5	&	3.57(-5)	&	$2.1 \times 1.7$	&	T15	&	2.30(-2)	&	2.50(-2)	\\
\hline																	
SiO	&	$5-4$	&	217.104980	&	31.3	&	5.21(-4)	&	$1.1 \times 0.8$	&	CALYPSO	&	9.58(-2)	&	1.20(-1)	\\
CH$_3$OH	&	$3_{1,3}-2_{1,2}$, A	&	143.865795	&	28.3	&	1.07(-5)	&	$2.2 \times 1.8$	&	T15	&	9.60(-2)	&	4.10(-2)	\\
NH$_3$	&	3, 3	&	23.870129	&	123.5	&	2.58(-7)	&	$2.0 \times 2.0$	&	C07	&	3.10(-3)	&	2.80(-3)	\\
\hline                  
\end{tabular}
\tablebib{
SOLIS-S1, -S3, and -S5 stand for Setups 1, 3, and 5 of the SOLIS data, respectively (see text for more details). C07, P12, and T15 stand for \citet{Choi2007}, \citet{Persson2012}, and \citet{Taquet2015}, respectively.
}
\label{transitions}
\end{table*}

%
For the SOLIS data, the IRAS4A binary system was observed in three different Setups (1, 3, and 5) with the IRAM NOEMA interferometer during several tracks in June and September 2016 \citep[see Table 1 of][]{Ceccarelli2017}. 
Setup 1, 3, and 5 cover the frequency ranges from 80.8 GHz to 84.4 GHz, from 95.7 GHz to 99.5 GHz, and between 204.0 and 207.6 GHz, respectively. 
For Setups 1 and 3, the array was used in both configurations D and C with baselines from 16 m to 240 m for Setup 1 and from 15 m to 304 m for Setup 3. For Setup 5, the array was used in configurations A and C.
The phase centre is on the IRAS4A1 source, 
$\alpha({\rm J2000})$ = 03$^h$ 29$^m$ 10$\fs$5, $\delta({\rm J2000})$ = +31$\degr$ 13$\arcmin$ 30$\farcs$9.
The bandpass was calibrated on 3C454.3 and 3C84, the absolute flux was fixed by observing MWC349, LKHA101, 2013+370, and 2007+659, and 0333+321 was used to set the gains in phase and amplitude.
The final uncertainty on the absolute flux scale is $\leq$15\%.
The phase rms was $\leq$50$\degr$, the typical precipitable water vapour (pwv) 
was from 4 mm to 15 mm, and the system temperatures ranged typically between 50 and 200 K.
The data were reduced using the packages CLIC and MAPPING of the GILDAS\footnote{http://www.iram.fr/IRAMFR/GILDAS/} software collection. 
A continuum map was obtained by averaging line-free channels and self calibrating the data. 
The self calibration solutions were then applied to the spectral cube, which was subsequently cleaned. 
The resulting synthesised FWHM beam is 4$\farcs$5$\times$3$\farcs$5 (P.A. =27$\degr$) for Setup 1, 2$\farcs$2$\times$1$\farcs$9 (P.A. =96$\degr$) for Setup 3, and 0$\farcs$9$\times$0$\farcs$7 (P.A. =47$\degr$) for Setup 5. 
The half power primary beam is 61$\farcs$4, 59$\farcs$2, and 24$\farcs$5 for Setups 1, 3, and 5 respectively.

We used the CALYPSO data observed with the IRAM-PdBI \citep{Maury2019} to extract the interferometric maps 
from the SiO($5-4$), SO($6_5-5_4$), and OCS($8-7$) transitions at $\sim 218$ GHz.  
These observations were carried out between February 2011
and February 2013 as part of the CALYPSO Large Program using the A and C 
configurations of the array. The synthesised FWHM beam at 1.4 mm is 1$\farcs$1$\times$0$\farcs$8. 
More details on the observational properties can be found in \citet{Santangelo2015}. 

IRAS4A was observed with the IRAM PdBI at 145 and 165 GHz on 2010 July 20, 21,
August 1, 3, November 24, and 2011 March 10 in the C and D most
compact configurations of the array. The baseline range of the
observations is 25 m - 140 m, allowing us to recover emission on scales
from 10-15$\arcsec$ to $\sim$ 2$\arcsec$. 
The WIDEX backends have been
used at 143.4 and 165.2 GHz, providing a bandwidth of 3.6 GHz each
with a spectral resolution of 1.95MHz ($\sim$ 3.5 – 4 km s$^{-1}$). High-resolution
narrowband backends focused on two CH$_3$OH lines have also been
used. They provide a bandwidth of 80 MHz with a spectral resolution of
0.04 MHz (0.08 km s$^{-1}$). We decreased the spectral resolution to 0.40 MHz
(0.8 km s$^{-1}$) to obtain a better signal-to-noise. 
More details on the observational dataset can be found in \citet{Taquet2015}.

PdBI observations at 225 GHz were performed on 27 and 28 November 2011,
on 12, 15, 21, 27 March 2012, and on 2 April 2012 in the B and C 
configurations of the array. The synthesised FWHM beam is 1$\farcs$2$\times$1$\farcs$0.
A more detailed description of the observational setups is provided by \citet{Persson2014}.

{ For all datasets, the amplitude calibration uncertainty is estimated to be $\sim15-20$ \%. 
Given we map relatively extended structures such as outflow lobes, we cleaned the images using the Hogbom method with natural weighting and with a mask containing all the molecular lobes to minimise artefacts in the cleaned image.}

\section{Results} \label{section3}

\subsection{Spatial distribution}

\begin{figure*}[htp]
\centering 
\includegraphics[width=1.705\columnwidth]{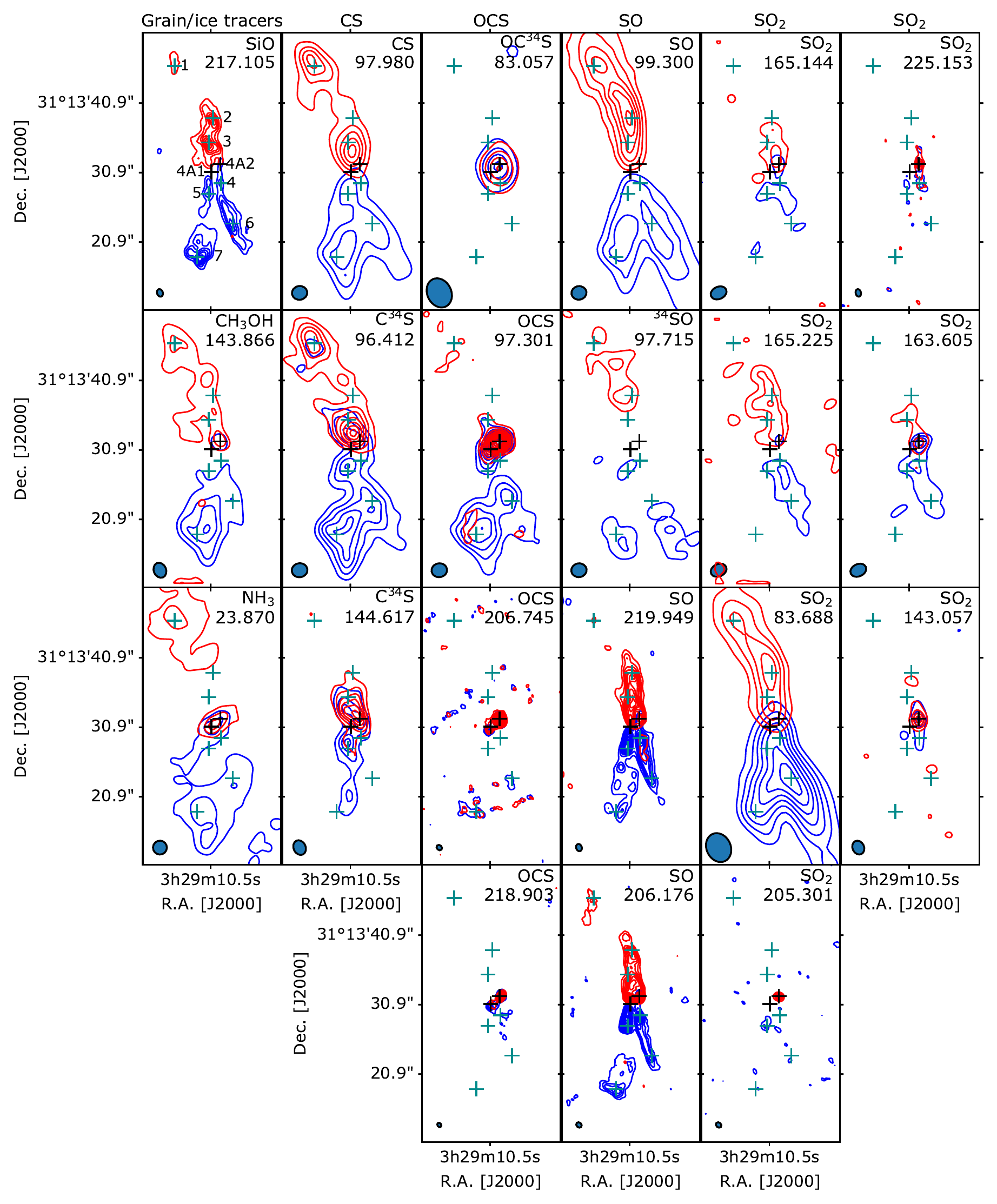}
\caption{Moment 0 integrated maps of the transitions from SiO, CH$_3$OH, NH$_3$, CS, OCS, SO, and SO$_2$ integrated over the (-10, +7) km s$^{-1}$ (blue) and (+7, 30) km s$^{-1}$ (red) ranges. Contour levels increase in steps of 3$\sigma$, the $\sigma$ noise levels of each map are listed in Table \ref{transitions}. { Transition frequency (in GHz) is given at the top right of each panel.} Black crosses represent the position of the 4A1 and 4A2 sources of the IRAS4A binary system. Green crosses represent the position of the SiO(5-4) emission peaks along the outflows. } 
\label{moment0_all}
\end{figure*}

\begin{figure*}[htp]
\centering 
\includegraphics[width=1.2\columnwidth]{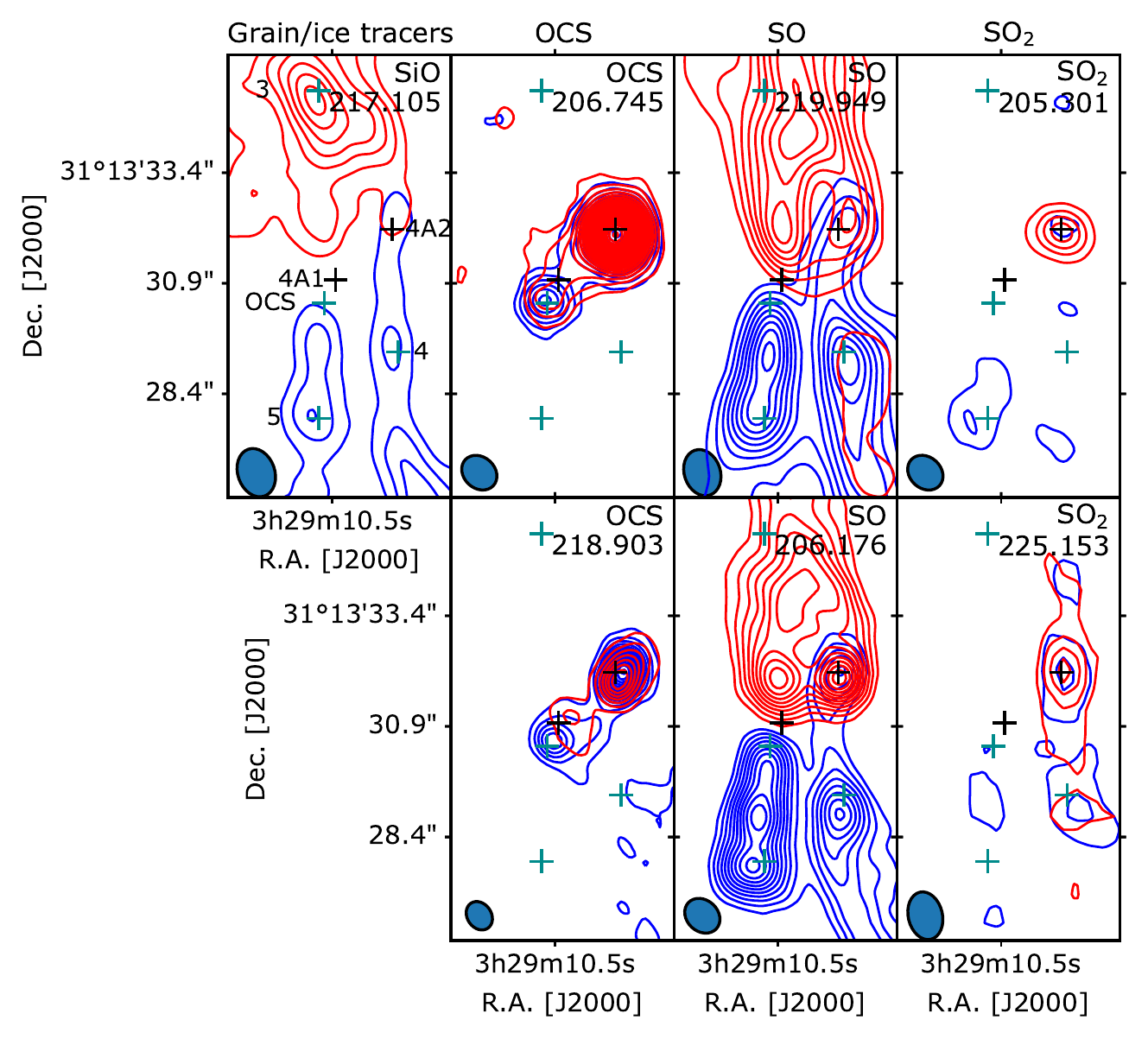}
\caption{Moment 0 integrated maps zoomed around the IRAS4A protostellar system of the transitions from SiO, OCS, SO, and SO$_2$ observed at high ($\sim 1 \arcsec$) angular resolution integrated over the (-10, +7) km s$^{-1}$ (blue) and (+7, 30) km s$^{-1}$ (red) ranges. Contour levels increase in steps of 3$\sigma$, the $\sigma$ noise levels of each map are listed in Table \ref{transitions}. { Transition frequency (in GHz) is given at the top right of each panel.} Black crosses represent the position of the 4A1 and 4A2 sources of the IRAS4A binary system. Green crosses represent the position of the SiO(5-4) and the blue OCS emission peaks along the outflows. } 
\label{moment0_highresol}
\end{figure*}

Integrated intensity maps from various sulfuretted transitions observed at 1\arcsec (hereafter called high) and about at 2-4\arcsec (hereafter called low) angular resolutions are shown in Figure \ref{moment0_all}. 'Zoomed' maps around the IRAS4A protobinary system of the transitions observed at 'high' angular resolution are also shown in Figure \ref{moment0_highresol}.
For comparison, we also show the emission from SiO($5-4$), NH$_3$(3, 3), and A-CH$_3$OH($3_{1,3}-2_{1,2}$) transitions. The SiO($5-4$) has been observed as part of the CALYPSO Large Program at $\sim$1\arcsec~resolution and was presented in \citet{Santangelo2015}. The A-CH$_3$OH($3_{1,3}-2_{1,2}$) transition at 143.866 GHz observed at $\sim$2$\arcsec$ was presented in \citet{Taquet2015}. A more detailed analysis of the CH$_3$OH emission along the outflows driven by IRAS4A is presented in a complementary study \citep{DeSimone2019}. The NH$_3$(3, 3) emission was observed by \citet{Choi2007} with the VLA interferometer with a $\sim$2\arcsec resolution.
SiO is thought to be formed in the gas phase from Si atoms initially locked at $\sim$90 \% in refractory grains and then sputtered in shocks whilst methanol CH$_3$OH and ammonia NH$_3$ are two abundant ice components. Methanol in particular is only formed at the surface of interstellar ices and cannot form efficiently through gas phase chemistry \citep{Watanabe2002, Geppert2006}. Since methanol should be efficiently destroyed in the hot ($T > 2000$ K) gas, it would therefore indicate the location of shock fronts where ice has been recently sputtered. SiO, NH$_3$, and CH$_3$OH can be therefore considered as reference species to locate the position of shocks throughout the outflows. The positions of the SiO($5-4$) emission peaks along the outflows are depicted by the green crosses with associated peak numbers. 

It can be clearly noticed that OCS, SO, and SO$_2$ targeted at high angular resolution trace different regions of the protostellar outflows. 
The blue emission of the SiO($5-4$) and the two SO transitions clearly allows us to disentangle the outflow emission from the two sources IRAS4A1 and A2. The outflow driven by IRAS4A1 seems to extend along the south-north (hereafter S-N) direction whilst the outflow driven by IRAS4A2 extends along the south west - north east (hereafter SW-NE) direction.
The SO($4_5-3_4$) transition at 206.176 GHz of our SOLIS data observed at high angular resolution therefore confirms the CALYPSO results already shown by \citet{Santangelo2015}. 

Unlike SO, OCS shows its brightest emission towards the hot corino of IRAS4A2. Interestingly, another OCS emission peaking at +3 km s$^{-1}$ is also detected at (+0.2$\arcsec$, -0.4$\arcsec$) with respect to 4A1 for the two targeted transitions and its position is depicted in the SiO map in Fig. \ref{moment0_highresol}. The location of this second OCS peak corresponds to the beginning of the south outflow driven by 4A1 detected in SO and SiO. 
The SO$_2$ transition shows extended emission but mostly in the blue range, around SiO peak 5 for the $11_{2,10}-11_{1,11}$ transition at 225.301 GHz and around SiO peak 4 for the $13_{2,12}-13_{1,13}$ one at 205.300 GHz. 

At large scales, extended emission from OCS is only detected in the blue (i.e. $V <7$ km s$^{-1}$) component, with a bright emission peak in the south outflow around SiO peak 7 driven by source 4A1 and a second fainter emission peak near SiO peak 6 along the NW-SE outflow.
SO and CS show extended emission in the blue and red components along the SW-NE and the south outflow with relatively similar intensities. For SO, the red component emission peaks follow SiO peaks 1, 2, and 3 whilst the blue component peaks are located near SiO peaks 5, 6, and 7. CS, on the other hand, does not show any bright emission near SiO peak 2.
SO$_2$ is brighter along the SW-NE outflow. In fact, SO$_2$ transitions at 143, 165, and 225 GHz are only detected along the SW-NE outflow, but the $8_{1,7} - 8_{0,8}$ transition at 83.688 GHz also extends towards the south lobe driven by 4A1. 

{ In order to compare the spatial distribution of the targeted molecules, we list in Table \ref{correlation_lowres} the Spearman\footnote{The Spearman correlation coefficient is defined as the Pearson correlation coefficient between the rank variables, they are consequently less affected by outliers.} ranking correlation coefficients of the blue and red emissions from the sulfuretted and methanol transitions observed at "low" angular resolution. To compute the "blue" ("red") coefficients between two given transitions, we included pixels where the "blue" ("red") integrated emissions of both transitions are higher than the 4$\sigma$ noise level. We verified that modifying the detection threshold between 3$\sigma$ and 5$\sigma$ for including the pixels in the coefficient estimation induces a variation of the Spearman coefficients by typically 20\% at most.
Sets of transitions with Spearman coefficients higher than 0.6 can be considered as strongly correlated whilst transitions with Spearman coefficients lower than 0.4 are considered as uncorrelated.
Since the "red" emission of the OCS transition is not detected along outflows, the corresponding Spearman coefficients are not included. 
}
As previously noted at first glance, the "blue" emission of OCS does not show any correlation with other sulfuretted species, with coefficients lower than 0.3, but seems to correlate relatively well with CH$_3$OH. SO$_2$ and SO seem to strongly correlate together with a coefficient higher than 0.8 for both emissions. The "blue" emission from CS seems to correlate well with SO and SO$_2$, with a coefficient higher than 0.6, confirming our first visual impression. However, the coefficients between the "red" CS emission and the "red" SO and SO$_2$ emissions are lower than 0.25. Indeed, CS is only bright near SiO peaks 1 and 3 whereas SO and SO$_2$ emissions also peak near SiO peak 2. 

To summarise, OCS, CS, SO, and SO$_2$ clearly show different spatial emission along the two outflows driven by IRAS4A. 
On large scales, OCS is brighter in the south outflow driven by IRAS4A1 whereas SO$_2$ is rather detected along the outflow driven by IRAS4A2 extended along the NE-SW direction. CS and SO show extended emission along the two outflows.
On smaller scales, an OCS emission peak is detected { close to the launching region of} the south outflow driven by 4A1 detected in SO and SiO. 

\begin{table*}[htp]
\centering
\caption{Spearman's rank correlation coefficients between the "blue" and "red" integrated emissions of molecules observed at low angular resolution.} 
\begin{tabular}{l c c c c c}
\hline              
\hline              
  & CH$_3$OH  & OCS & CS  & SO  & SO$_2$  \\
  & ($3_{1,3}-2_{1,2}$) & ($8-7$) & ($2-1$) & ($3_2 - 2_1$) & ($8_{1,7}-8_{0,8}$) \\  
  & Blue / Red & Blue / Red & Blue / Red & Blue / Red & Blue / Red \\      
\hline        
CH$_3$OH($3_{1,3}-2_{1,2}$) & 1 / 1         & 0.61 / -~~~~~  & 0.77 / 0.26 & 0.34 / 0.20 & -0.08 / 0.21~ \\
OCS($8-7$)                  & 0.61 / -~~~~~ & 1 / 1          & 0.27 / -~~~~~ & 0.09 / -~~~~~ & 0.14 / --~~~~~     \\
CS($2-1$)                   & 0.77 / 0.26   & 0.27 / -~~~~~  & 1 / 1       & 0.66 / 0.01 & 0.66 / 0.24  \\
SO($3_2 - 2_1$)             & 0.34 / 0.20   & 0.09 / -~~~~~  & 0.66 / 0.01 & 1 / 1       & 0.83 / 0.84  \\
SO$_2$($8_{1,7}-8_{0,8}$)   & -0.08 / 0.21~ & 0.14 / -~~~~~  & 0.66 / 0.24 & 0.83 / 0.84 & 1 / 1        \\ 
\hline                  
\end{tabular}
\tablefoot{
Left and right values correspond to the Spearman's rank correlation coefficients obtained for the "blue" and "red" emissions, respectively, when the emissions from both considered transitions are higher than the 4$\sigma$ noise level.
}
\label{correlation_lowres}
\end{table*}

\subsection{Opacity estimates} \label{opacity}

\begin{figure}[htp]
\centering 
\includegraphics[width=0.26\textwidth]{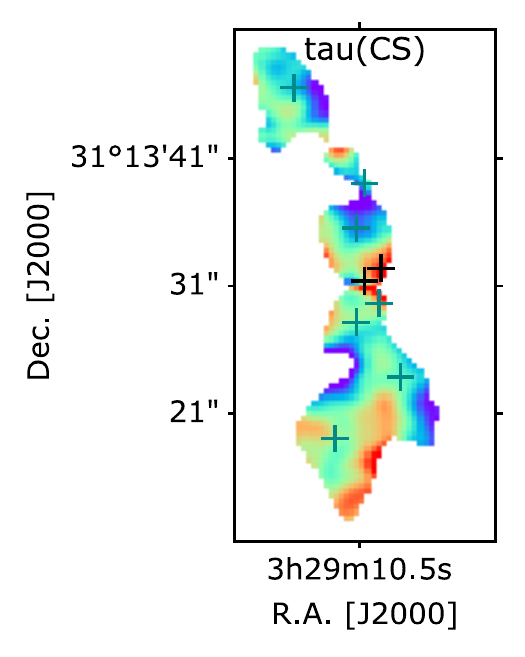}
\includegraphics[width=0.218\textwidth]{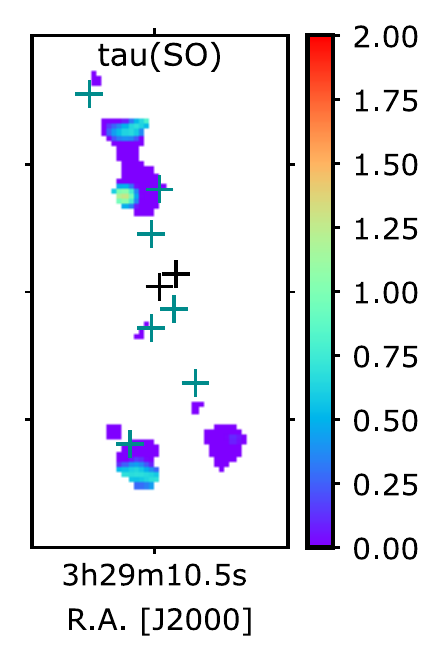}
\caption{Opacity maps of the CS($2-1$) and SO($3_2-2_1$) transitions estimated from the intensity ratio of the $^{32}$S and $^{34}$S transitions when the flux of both isotopologues is higher than four times the $\sigma$ rms noise level and assuming the the $^{34}$S transitions are optically thin (see text for more details). Black crosses represent the position of the 4A1 and 4A2 sources of the IRAS4A binary system. Green crosses represent the position of the SiO(5-4) emission peaks along the outflows. } 
\label{opacity_maps}
\end{figure}

We detected extended emission of the CS($2-1$) and SO($3_2-2_1$) transitions from the two $^{32}$S and $^{34}$S isotopologues with similar observational properties. The emission maps therefore allow us to estimate the opacities of the main $^{32}$S transitions, $\tau_{32}$, towards each pixel of the map. 
Assuming optically thin emission from the $^{34}$S isotopologues, a Local Thermodynamic Equilibrium (LTE) population of the levels, and same rotational temperature $T_{\rm rot}$ for both isotopologues, $\tau_{32}$ can be derived from the following formula \citep[see][for more details]{Goldsmith1999}
\begin{equation}
\begin{split}
\frac{\tau_{32}}{1-\exp(-\tau_{32})} = & \left(\frac{N_{32}}{N_{34}}\right) \times \left(\frac{Z_{34}}{Z_{32}}\right) \times \left(\frac{\nu_{34}}{\nu_{32}}\right)^2 \times \left(\frac{A_{u,l,32}}{A_{u,l,34}}\right) \times \left(\frac{g_{up,32}}{g_{up,34}}\right) \times  \\
& \left(\frac{W_{34}}{W_{32}}\right) \times \exp{\left(\frac{E_{up,34}-E_{up,32}}{k T_{rot}}\right)} 
\end{split}
\label{eq_tau}
\end{equation}
where  $\left(\frac{N_{32}}{N_{34}}\right)$ is the column density ratio between the $^{32}$S and $^{34}$S isotopologues of each species, assumed to be equal to the elemental $^{32}$S/$^{34}$S ratio of 22 in the local interstellar medium \citep{Wilson1994};
$Z$, $\nu$, $A_{u,l}$, $g_{up}$, $W$, and $E_{up}$ are the partition functions at a given $T_{rot}$, the frequency, the Einstein-A coefficient, the statistical weight, the integrated brightness temperature, and the upper level energy of the ($^{32}$S or $^{34}$S) transition, respectively. We then numerically interpolated the opacity $\tau_{32}$ from the left term of equation \ref{eq_tau} towards each pixel of the map if the intensity of the $^{32}$S and $^{34}$S are both higher than the 4$\sigma$ noise level.

Figure \ref{opacity_maps} shows the opacity maps of the CS($2-1$) and SO($3_2-2_1$) transitions for $T_{rot} = 30$ K. It should be noted that the results do not significantly depend on the assumed value of $T_{rot}$ as long as $T_{rot}$ is higher than 10 K, due to the similar upper level energies between the transitions of the $^{32}$S and $^{34}$S isotopologues (differences of 0.9 and 0.1 K for CS and SO, respectively). { Opacity errors induced by the flux calibration uncertainties are about 30\%.}
It can be seen that the CS($2-1$) transition is moderately optically thick with $\tau \sim 1$ throughout most of the outflows and shows an opacity higher than 2 close to source 4A2. In contrast, SO($3_2-2_1$) is optically thin with $\tau << 1$ in most of the outflows.

\subsection{Kinematics along the outflows}

\begin{figure*}[htp]
\centering 
\includegraphics[width=\columnwidth,valign=t]{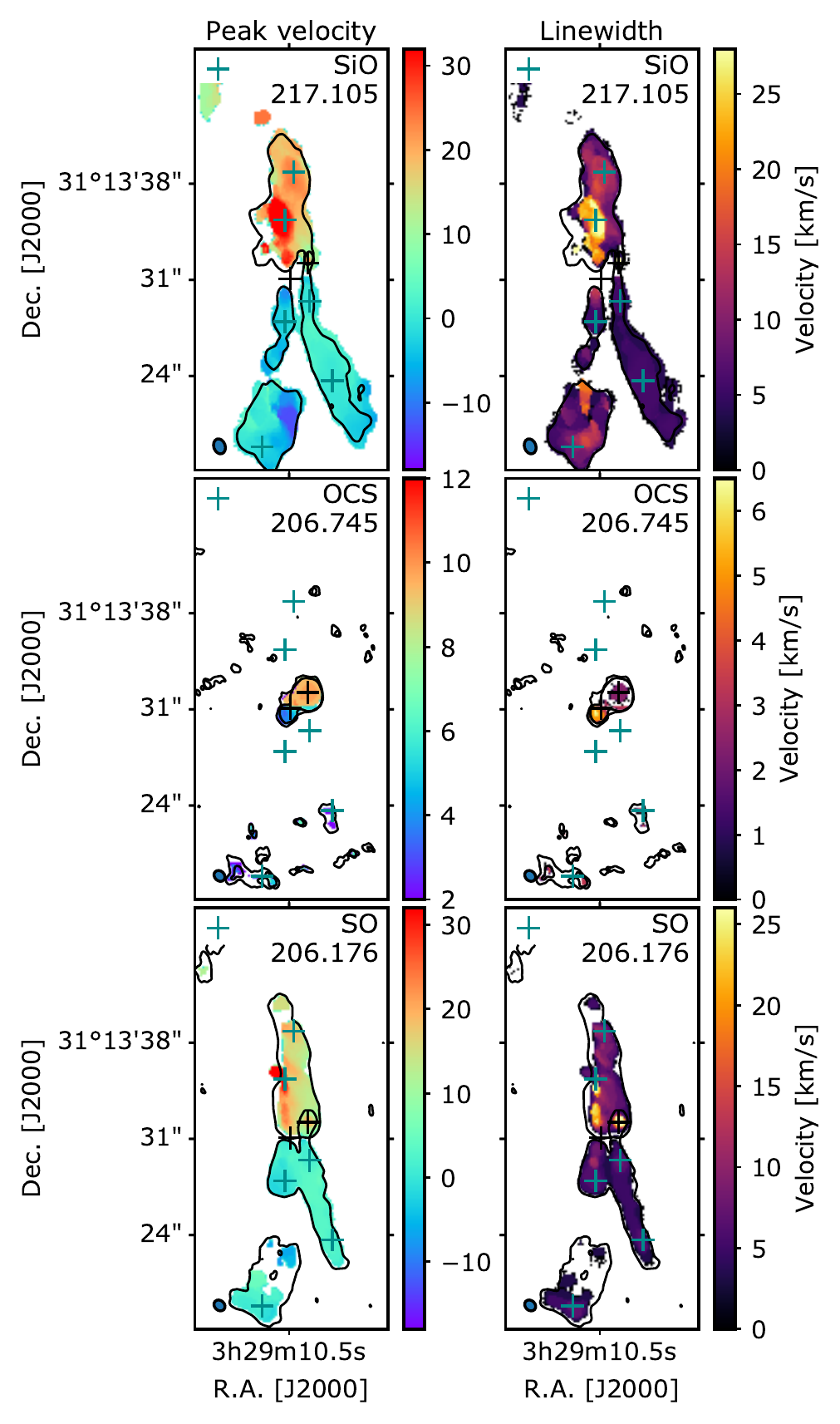} 
\includegraphics[width=\columnwidth,valign=t]{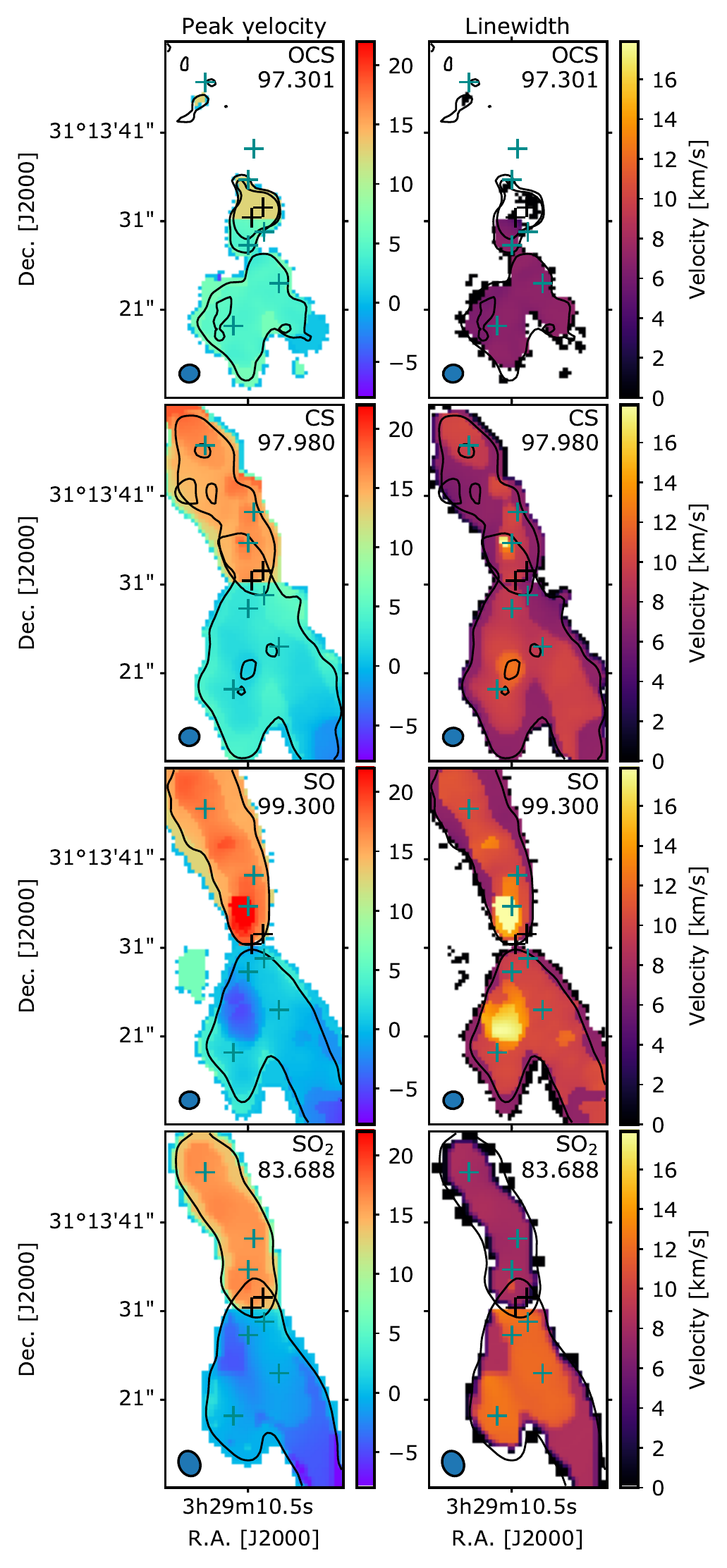} 
\caption{Moment 1 and moment 2 maps of the SiO($5-4$), OCS($17-16$), and SO($6_5-5_4$) emission observed at "high", 1 $\arcsec$, angular resolution (left panels) and of the OCS($8-7$), SO($3_2-2_1$), CS($2-1$), and SO$_2$($8_{1,7}-8_{0,8}$) emission at "low", $2-4$ \arcsec, resolution (right panels).
Contours represent the 3$\sigma$ level. Black crosses represent the position of the 4A1 and 4A2 sources of the IRAS4A binary system. Green crosses represent the position of the SiO(5-4) emission peaks along the outflows. 
} 
\label{mommaps}
\end{figure*}

Figure \ref{mommaps} presents the moment 1 and moment 2 maps of SiO and the sulfuretted transitions observed at "high" and "low" angular resolutions. As supporting material, Figures \ref{channelmaps_lowresol} and \ref{channelmaps_highresol} present the velocity emission maps whilst Figures \ref{spectra_sionorth} and \ref{spectra_siosouth} of the Appendix show the spectra of the sulfuretted species obtained towards the SiO and the OCS emission peaks depicted in the SiO maps in Fig. \ref{moment0_all} and \ref{moment0_highresol}. 

The moment 1 and moment 2 maps at "high" angular resolution of SiO and SO clearly allow us to disentangle the two components of the molecular outflows both in the north and the south parts. 
In the northern part, the outflow driven by 4A1 is much faster with a velocity peak of $\sim 30$ km s$^{-1}$ and a velocity linewidth of $\sim 25$ km s$^{-1}$, both seen in SiO and SO. On the other hand, the outflow driven by IRAS4A2 seems to be slower, with a peak velocity of $\sim 20$ km s$^{-1}$ and a velocity linewidth of $\sim 10$ km s$^{-1}$ for both molecules. 
In the southern part, the S-N outflow driven by 4A1 is slightly faster with a peak velocity of $\sim -5$ km s$^{-1}$ (or $\sim 12$ km s$^{-1}$ relative to the source velocity) whereas the SW outflow driven by 4A2 has a peak velocity of $\sim 0 - 5$ km s$^{-1}$. 

The low angular resolution maps are obtained at lower frequency, between 80 and 100 GHz, decreasing the velocity resolution by a factor of $\sim 2.5$ with respect to the high angular resolution maps at $215-220$ GHz ($\sim 6$ vs $\sim 2.5$ km s$^{-1}$).
Nevertheless, the low angular resolution moment maps of SO($3_2-2_1$) also reveal high velocity emission associated with the S-N outflow driven by 4A1, with peak velocities of +20 and -5 km s$^{-1}$ in the north and south, respectively, and velocity linewidth of 15 km s$^{-1}$, confirming the high angular resolution maps. The low velocity resolution of the low-angular resolution maps prevents us to detect any weaker evolution of the velocity along the outflows for the other species.

\subsection{Comparison with the IRAM-30m spectra} \label{IRAM30mspectra}

In order to estimate the flux resolved out by the NOEMA interferometer, we compared the spectra obtained with our NOEMA datasets with the single-dish spectra obtained with the IRAM-30m single-dish telescope as part of the ASAI Large Program \citep{Lefloch2018}. Briefly, the aim of the ASAI program was to carry out an unbiased spectral survey at high spectral resolution of a sample of young nearby low-mass and intermediate-mass protostars in the { 3, 2, and 1.3 mm} spectral windows. 
To obtain the NOEMA spectra, we estimated the flux of each spectral channel coming from a circular mask of a size equal to the beam of the IRAM-30m telescope that ranges from $\sim 30 \arcsec$ at 80 GHz to $\sim 10 \arcsec$ at 220 GHz.

\begin{figure}[htp]
\centering 
\includegraphics[width=\columnwidth]{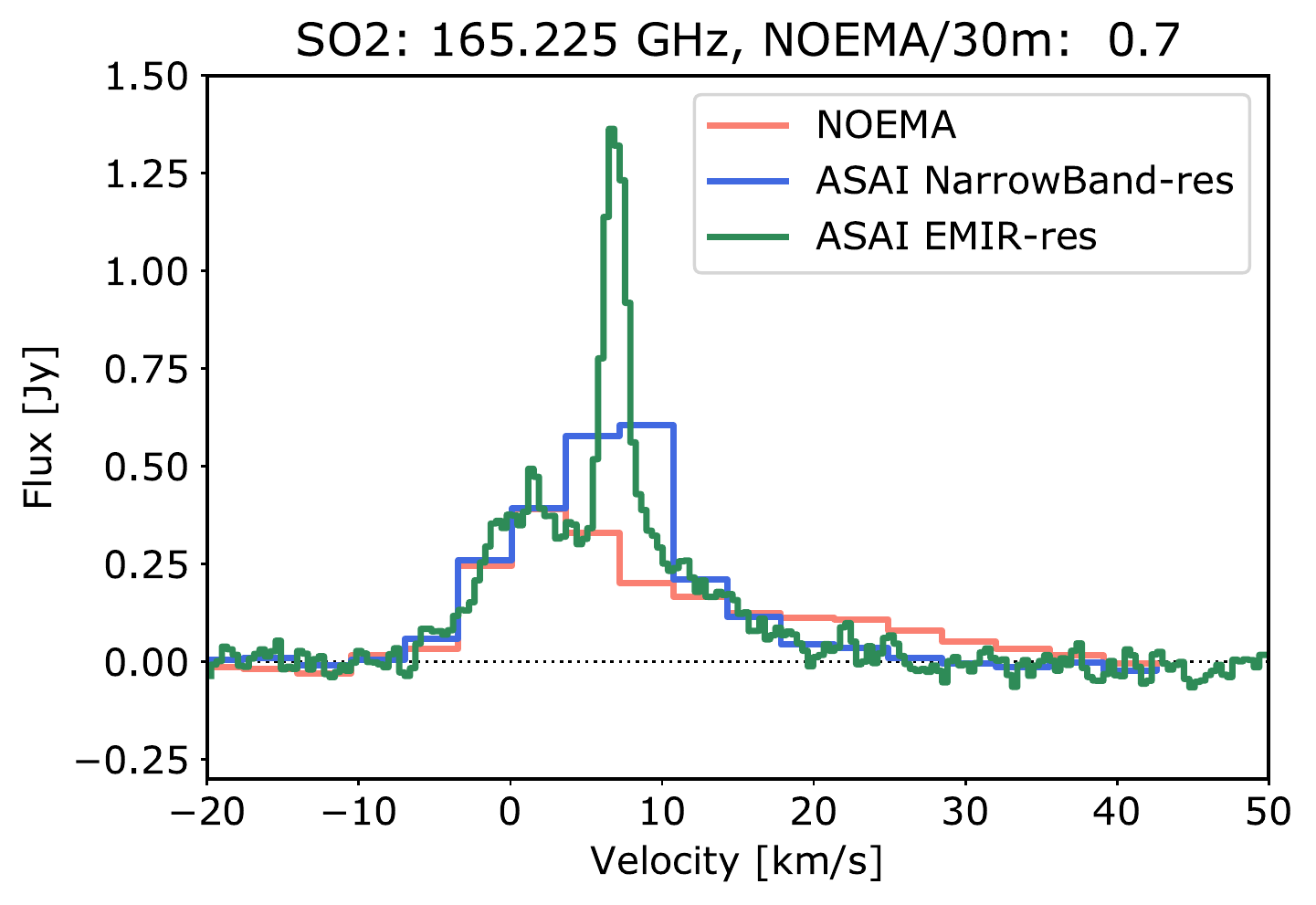}
\caption{Flux spectra of SO$_2$ at 165.225 GHz observed with the IRAM-30m with the ASAI Large Program with the EMIR spectral resolution (in green) and smoothed over the WIDEX resolution (in blue) and with the NOEMA interferometer assuming a mask equal to the IRAM-30m beam (in red). The flux recovered by NOEMA is reported on top of the panel (see text in section \ref{IRAM30mspectra} for more details). } 
\label{spectra_noema_30m_so2}
\end{figure}

The single-dish spectra of transitions from sulfuretted molecules, SiO, and CH$_3$OH detected in the ASAI data are compared with the NOEMA spectra in Figure \ref{spectra_noema_30m} of the Appendix. { We also show one representative transition is shown in Fig. \ref{spectra_noema_30m_so2}, the SO$_2$ transition at 165.225 GHz.} We present the ASAI spectra both for the original 200 kHz spectral resolution obtained with the EMIR receiver connected to the FTS backend and for a smoother spectral resolution of 1.95 MHz equivalent to the WIDEX correlator of NOEMA. 
The recovered flux of transitions detected in the ASAI data at 5$\sigma$ are estimated from the ratio of the flux integrated between -20 and +50 km s$^{-1}$ and are given at the top of each panel. 

High-resolution spectra from ASAI can usually be decomposed into several components \citep[see for instance][for a more detailed description of a similar dataset towards an intermediate-mass protostar]{Ospina-Zamudio2018}: 
1) a narrow line component associated with the large scale cold and quiescent envelope, 
2) a broad line component associated with the outflowing gas as seen with broad wings, 
3) a moderately-broad line component associated with the hot corino. 
Transitions with a low upper level energy, such as the OCS($8-7$), SO($3_2-2_1$), SO($6_5-5_4$), SO($4_5 - 3_4$), SO$_2$($8_{1,7}-8_{0,8}$), or SO$_2$($7_{1,7}-6_{0,6}$) show a bright narrow line component observed with the IRAM-30m but not detected with NOEMA. This suggests that the cold extended emission is completely resolved out by the NOEMA interferometer with the used configurations of the array. 
The velocity profiles of the moderately-broad and broad components for the SO, SO$_2$, CH$_3$OH transitions show a very good agreement between the IRAM-30m and NOEMA datasets, confirming that the high-velocity gas is associated with compact emission. 
The fluxes recovered by NOEMA typically range from $\sim 25-50$ \% for transitions showing a bright narrow line component to $\sim 100$ \% ($\pm 20$\% of calibration and statistical errors) for species mostly associated with compact gas.

\subsection{Dissecting the molecular outflows with clustering methods}

The independent analysis of the moment 0 maps, namely the integrated intensity maps of the molecular emission, and then the kinematics information included in the moment 1 and 2 maps already allowed us to distinguish by eye several components along the two molecular outflows driven by the IRAS4A protobinary system. 
We now aim to combine the information included in the three types of moment maps in order to quantitatively distinguish the spatial components included in the outflows through the use of a clustering method. Among the variety of clustering methods found in the literature, we chose the relatively simple and commonly used K-means algorithm \citep{Lloyd1982} available within the \texttt{scikit-learn} Python package.
Our data are a set of pixels each characterised by two sky coordinates and $3 \times N_{\rm species}$ (integrated intensity, velocity peak, and linewidth for each species) features. The clustering algorithm only uses the molecular emission information for clustering the outflows, ignoring their spatial coordinates. { Before performing the clustering analysis, we normalised the data so all features have the same mean and standard deviation. }
Briefly, K-means, also called Lloyd's algorithm, tries to decompose a set of $N$ samples (i.e. here the pixels) into $K$ disjoint clusters each described by a mean, also called centroid, by minimising their inertia, or within-cluster sum of criterion. In practice, the K-means algorithm has three steps:
1) K-Means first chooses the initial centroids. Here we employed the so-called "k-means++" initialisation scheme \citep{David2007} in order to chose initial centroids that are distant from each other.
2) The algorithm then assigns each sample to its nearest centroid, namely the one with the shortest distance.
3) K-Means then creates new centroids by taken the mean value of all of the samples assigned to each previous centroid. The difference between the old and the new centroids are computed and the algorithm repeats the last two steps until this value is less than a given threshold. 

\begin{figure}[htp]
\centering 
\includegraphics[width=0.8\columnwidth]{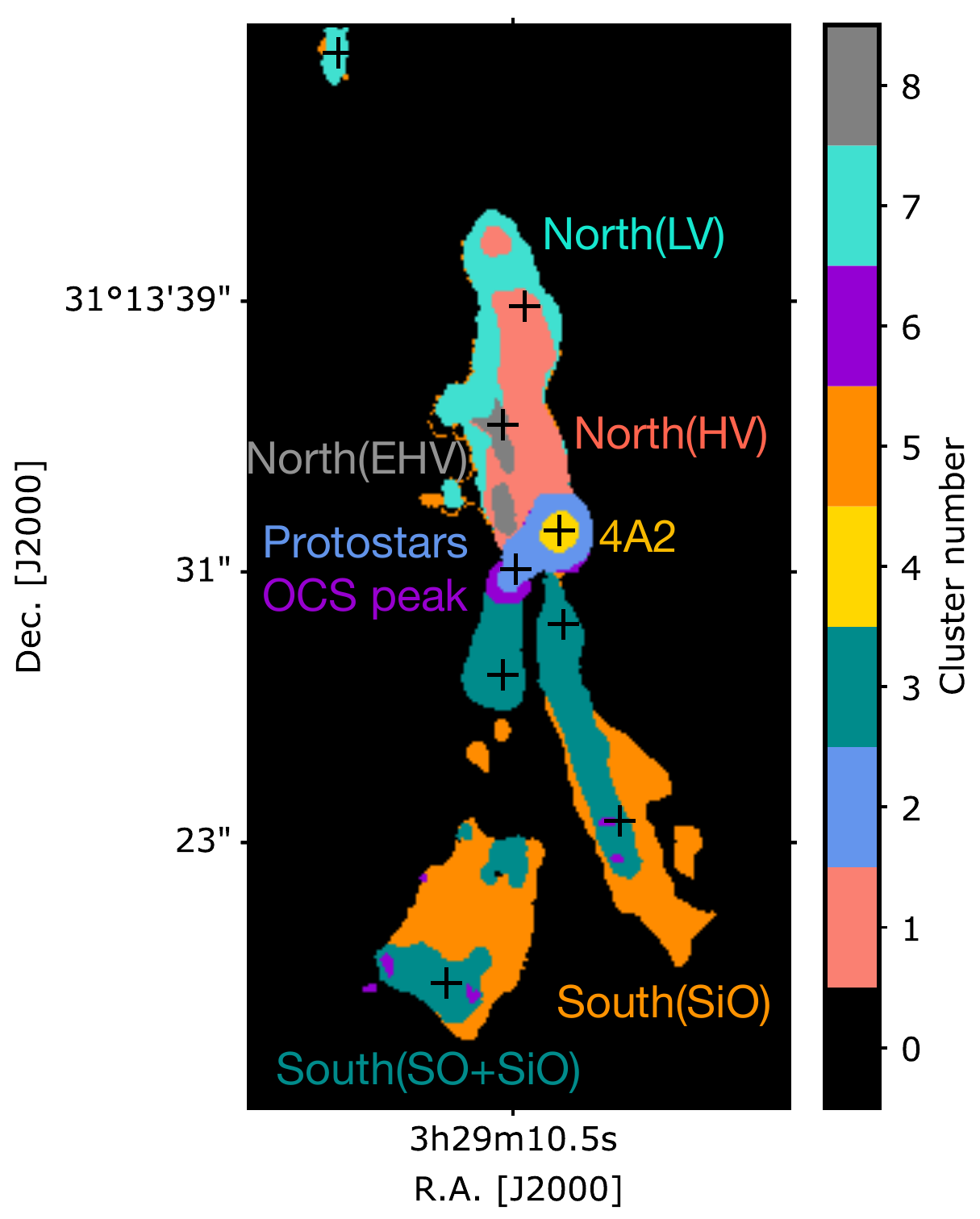}
\caption{Map of the clusters estimated by the K-Means algorithm based on the moment 0, 1, and 2 maps of the SiO($5-4$), SO($4_5-3_4$), and OCS($17-16$) molecular emissions (see text for more details).} 
\label{clusters_highres}
\end{figure}

We tried to apply this clustering method to the SiO($5-4$), SO($4_5-3_4$), and OCS($17-16$) molecular emission observed at high angular resolution, and to the OCS($8-7$), SO($3_2-2_1$), CS($2-1$), and SO$_2$($8_{1,7}-8_{0,8}$) emission at "low" resolution. However, the velocity resolution of the low-angular resolution maps { is relatively low ($\sim 6-7$ km s$^{-1}$) resulting in only a few discrete velocity peak and linewidth values that cannot be quantitatively used for correctly clustering the outflows}. Figure \ref{clusters_highres} therefore only shows the map of clusters identified from the high angular resolution maps.  
%
The number of clusters is given as input parameter. 
We increased the number of clusters from 3 to 9 (i.e. the number of features) and identify gradually the components in the corresponding cluster maps. 
For three clusters (from cluster 0 to 2 in Fig. \ref{clusters_highres}), the noise in black, the north outflows seen in SO and SiO, and the central "Protostars" component seen in OCS in blue are identified. 
For four clusters, the south outflow component is identified.  
For five clusters, the hot corino from 4A2 is identified in yellow. 
For six clusters, the north outflow is decomposed into two low-velocity ("North(LV)" in cyan) and high-velocity ("North(HV)" in salmon) components.
For seven clusters, the south outflow is decomposed into a SiO-only emission ("South(SiO)" in orange) and SO+SiO emission ("South(SO+SiO)" in green).
For eight clusters, the OCS peak (in purple) is identified.
For nine clusters, the extremely-high-velocity component ("North(EHV)" in grey) is clearly identified, as already shown by the moments 1 and 2 in Fig. \ref{mommaps}. 

We also estimated the effect of noise on the clustering results by adding a random Gaussian noise to the moment 0 maps with standard deviations equal to those listed in Table \ref{transitions}. Adding a random noise to our dataset gives very similar results. All clusters but the North(EHV) are identified, whilst cluster 0 depicting the noise in black is decomposed into two randomly spatially distributed clusters.

This simple clustering method therefore allows us to quantitatively dissect the outflows into several components that were already qualitatively suspected by eye. More sophisticated clustering methods have been proven to be efficient to dissect larger and more complex regions where a visual inspection is impossible \citep[see][for instance]{Bron2018}.

\section{Radiative transfer analysis}

\subsection{Rotational diagram analysis}

The moment maps from the SO$_2$ transitions at $\sim 143$ and $\sim 165$ GHz share similar observational properties (similar frequency, angular resolution, configuration of the interferometer array, and rms noise). Their detection therefore allows us to self consistently analyse their emission and therefore to investigate the evolution of the physical conditions along the outflows driven by IRAS4A.
We start with a Rotational Diagram (RD) analysis to estimate the SO$_2$ column density and the rotational temperature throughout the outflows \citep{Goldsmith1999}. For this purpose, we extracted the flux of each pixel, set to 0.3 arcsec, of the velocity-integrated blue-shifted (-10; +7 km s$^{-1}$) and red-shifted (+7; +20 km s$^{-1}$) interferometric maps for the four SO$_2$ transitions at $\sim 143$ and $\sim 165$ GHz with Einstein coefficients higher than $10^{-5}$ s$^{-1}$, spanning a range of upper level energies between 23 and 137 K. 
The RD analysis is performed assuming Local Thermal Equilibrium (LTE) and optically thin emission when at least three SO$_2$ transitions show a flux higher than 4	$\sigma$, $\sigma$ being the noise level in each moment 0 map listed in Table \ref{transitions}. 

\begin{figure}[htp]
\centering 
\includegraphics[width=\columnwidth]{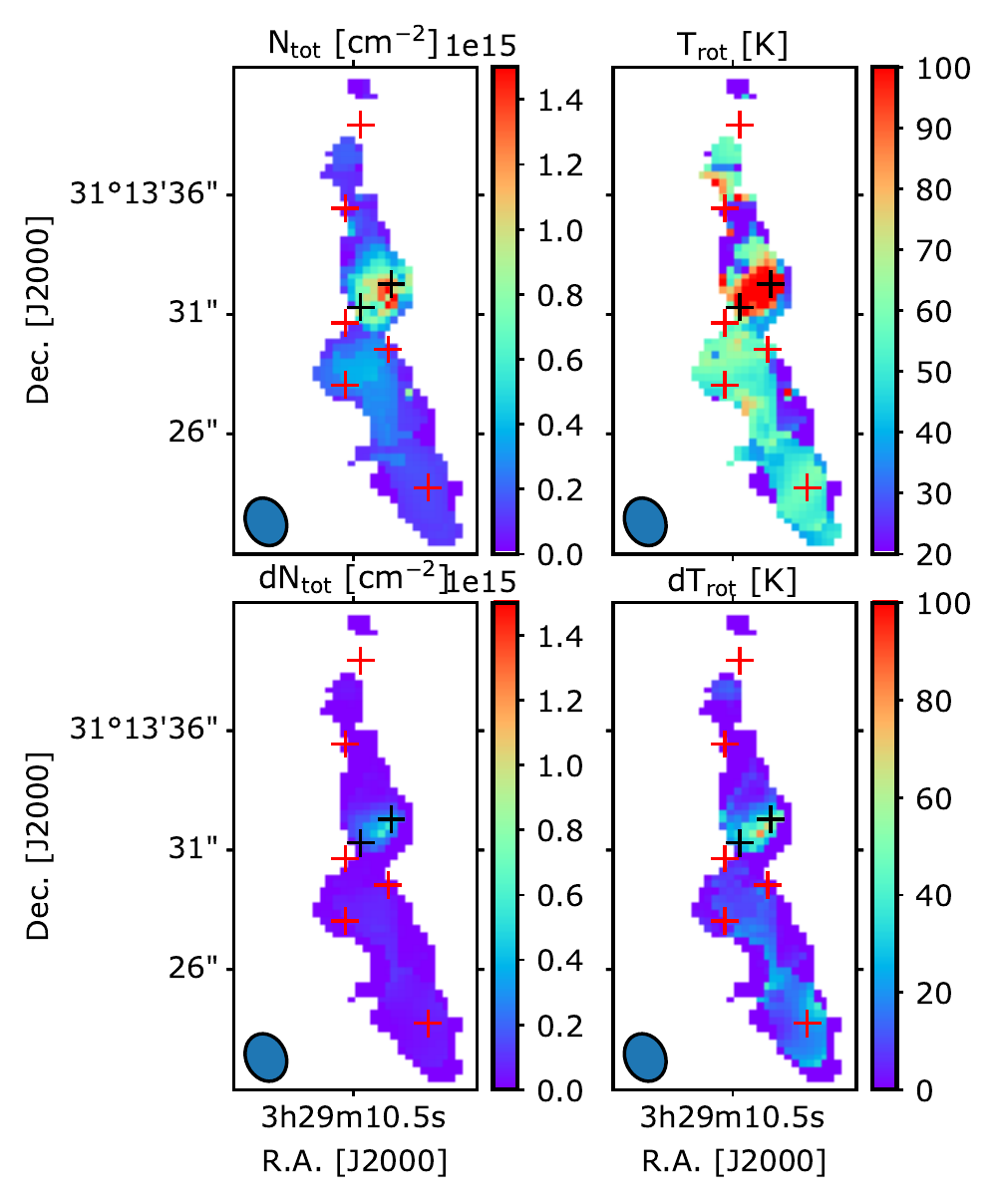}
\caption{Column density $N_{\rm tot}$ and rotational temperature $T_{\rm rot}$ and maps (top panels) together with their uncertainties $dN_{\rm tot}$ and $dT_{\rm rot}$ (bottom panels) obtained from a rotational diagram analysis when the flux of at least three SO$_2$ transitions is higher than four times the $\sigma$ rms noise level of each map. } 
\label{RD_maps}
\end{figure}

Figure \ref{RD_maps} shows the SO$_2$ rotational temperature and column density maps together with their uncertainties derived from the RD analysis. 
The SO$_2$ column density is about  $3 - 5 \times 10^{14}$ cm$^{-2}$ along the outflow driven by 4A2 and increases up to $1.4 \times 10^{15}$ cm$^{-2}$ near source 4A2. 
The rotational temperature of SO$_2$ remains close to $50-60$ K throughout the outflow and increases up to $\sim 100$ K in the hot corino of IRAS4A2. In order to assess whether the lower SO$_2$ rotational temperature derived in the outflow with respect to the hot corino could be due to sub-thermal excitation of the levels, given the high critical densities of its transitions ($1 - 3 \times 10^6$ cm$^{-3}$), or to lower kinetic temperatures, a non-LTE analysis taking the collisional and radiative processes into account is needed. 


For other species, the low number of transitions and their similar upper level energies prevent us to carry out a full rotational diagram analysis. Following the opacity estimates carried out in section \ref{opacity}, we estimated the column density ratios along the outflows using the OCS($8-7$), C$^{34}$S($2-1$), SO($3_2-2_1$), and SO$_2$($8_{1,7}-8_{0,8}$) transitions that have close upper level energies between 7 and 37 K. 
We assumed a rotational temperature $T_{\rm rot} = 50$ K following the SO$_2$ rotational diagram analysis and optically thin emission. 

\begin{figure*}[htp]
\centering 
\includegraphics[width=0.9\textwidth]{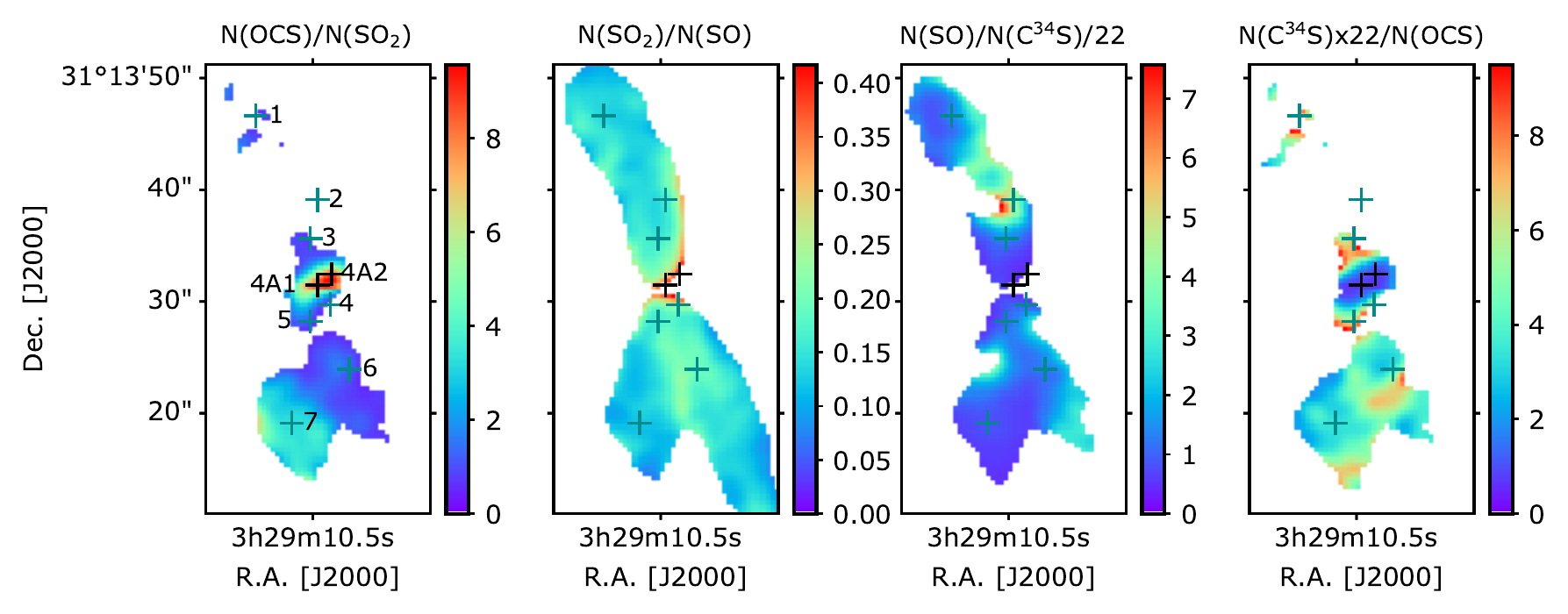}
\caption{Column density ratio maps betweeen OCS, SO$_2$, SO, and CS obtained from a rotational diagram analysis when the flux of of both transitions are higher than six times the $\sigma$ rms noise level. } 
\label{aburatio_maps}
\end{figure*}

Figure \ref{aburatio_maps} presents the column density ratio maps between OCS, SO$_2$, SO, and CS. As expected from the intensity maps, the OCS/SO$_2$ abundance ratio increases from $\sim 1$ in the SW outflow driven by 4A2 to $\sim 4$ in the south outflow driven by 4A1. The SO$_2$/SO abundance ratio slightly decreases from 0.2 to 0.1 between the SW outflow and the south outflow. One can notice that the SO abundance is remarkably high with respect to CS towards SiO peak 2, with a SO/CS abundance ratio of 7, whilst it remains close to 1 throughout the outflows. { Column density ratio errors induced by the flux calibration uncertainties are about 30\%.}

\subsection{Non-LTE RADEX Analysis} \label{radex}

A non-LTE analysis of the SO$_2$ emission, the only molecule studied in this work with at least four detected transitions with similar observational properties, is performed by using the non-LTE radiative transfer code RADEX \citep{vanderTak2007}.
To this mean, we assumed an isothermal and homogeneous medium without large-scale velocity fields, a comparable method to the Large Velocity Gradient (LVG) analysis. 
We used the collision rate coefficients of SO$_2$ with H$_2$ computed by \citet{Balanca2016} within the infinite order sudden (IOS) approximation in the 100-1000 K temperature range. The line excitation temperatures $T_{{\rm ex}}$, the opacities $\tau$, and the integrated brightness temperatures $T_{\rm mb}$ were computed by varying the SO$_2$ column density between $10^{13}$ and $10^{15}$ cm$^{-2}$, the H$_2$ density between $10^4$ and $10^8$ cm$^{-3}$, and the kinetic temperature $T_{\rm kin}$ between 50 and 550 K.   
We assumed a FWHM linewidth of 6 and 3 km s$^{-1}$ for the blue- and red-shifted emissions, respectively. 
The analysis was performed towards the SiO peaks where at least three of the 165.144, 165.225, 163.605, and 143.057 GHz SO$_2$ transitions, observed with similar observational properties, are detected at a 3$\sigma$ level. 
Best-fit properties together with their 1$\sigma$ uncertainties for the physical conditions and the excitation temperatures are given in Table \ref{LVGresults}. 

For the SiO peaks where the RADEX analysis has been performed, the SO$_2$ emission is reproduced for densities between $10^5$ and $10^6$ cm$^{-3}$. 
These best-fit densities are lower than the critical densities of all SO$_2$ transitions ($\sim 3 \times 10^6$ cm$^{-3}$ for the transitions at 163 - 165 GHz and $1 \times 10^6$ cm$^{-3}$ at 143.057 GHz). 
The three SiO peaks where the best-fit density is about $10^6$ cm$^{-3}$ show best-fit temperatures lower than 100 K whilst the two SiO peaks with a lower best-fit density of $2-3 \times 10^5$ cm$^{-3}$ show best-fit temperatures higher than 200 K. 
The different physical conditions inferred from the RADEX analysis could be due to different physical shock conditions or emitting SO$_2$ positions within shocks. But given the best-fit densities found here, it is possible that SO$_2$ traces the already compressed thermalisation zone or the post-shock region where densities are higher. 


All transitions but the one at 165.225 GHz show low excitation temperatures between 10 and 40 K. 
In contrast, the excitation temperature of the transition at 165.225 GHz is higher than the kinetic temperature towards SiO peaks 2, 3, and 5 whilst it is negative towards SiO peak 4. This suggests that this transition could be masering, explaining its brighter emission with respect to others with similar spectroscopic properties. 

%
Column densities derived with the non-LTE analysis are about $1.3 - 2.5 \times 10^{14}$ cm$^{-2}$ and are very similar to those derived with the RD analysis. They therefore suggest that the RD analysis gives a good estimate of the SO$_2$ column density throughout the outflow. 
For the best-fit properties, SO$_2$ transitions are optically thin with opacities lower than 0.1.

\begin{table*}[htp]
\centering
\caption{Best-fit models and 1$\sigma$ uncertainties of the non-LTE
  RADEX analysis of SO$_2$ emission towards several SiO peaks.} 
\begin{tabular}{l c c c c c c c c c}
\hline					
\hline					
Peak  & \multicolumn{2}{c}{Coordinates} & \multicolumn{3}{c}{Best-fit properties} & \multicolumn{4}{c}{$T_{\rm ex}$ (K)}  \\
\hline                                      
  & R.A.  & Dec.  & $n_{\rm H2}$  & $T_{\rm kin}$ & $N_{\rm tot}$ & 165.144 & 165.225 & 163.605 & 143.057 \\
  & Offset  & Offset  & (cm$^{-3}$) & (K) & (cm$^{-2}$) & (GHz) & (GHz) & (GHz) & (GHz) \\
\hline                                      
SiO 2 (4A2 outflow: N)  & -0.1  & 7.5 & $8^{+12}_{-5} \times 10^5$  & $95^{+40}_{-30}$  & $2.5 \pm 0.5 \times 10^{14}$  & 26  & 204 & 33  & 11  \\
SiO 3 (4A2 outflow: N)  & 0.5 & 4.1 & $1.2^{+5.0}_{-0.8} \times 10^6$ & $70^{+50}_{-20}$  & $2.5 \pm 1.0 \times 10^{14}$  & 38  & 115 & 39  & - \\
SiO 4 (4A2 outflow, SW)  & -1.3  & -1.8  & $3.2^{+4.0}_{-1.0} \times 10^5$ & $325^{+225}_{-75}$  & $1.5 \pm 0.5 \times 10^{14}$  & 24  & -112  & 26  & 9 \\
SiO 5 (4A1 outflow, SE)  & 0.5 & -3.2  & $1.2^{+3.0}_{-0.5} \times 10^6$ & $70^{+30}_{-10}$  & $2.5 \pm 0.5 \times 10^{14}$  & 38  & 115 & 39  & 14  \\
SiO 6 (4A2 outflow, SW)  & -2.9  & -7.5  & $1.8^{+4.0}_{-1.0} \times 10^5$ & $200^{+250}_{-80}$  & $1.3 \pm 0.5 \times 10^{14}$  & 10  & 48  & 20  & - \\
\hline									
\end{tabular}
\label{LVGresults}
\end{table*}


\section{Comparison with astrochemical predictions}


\subsection{The modelling}

In this section, we aim to interpret the chemical evolution of sulfuretted species observed along the outflows driven by the NGC1333-IRAS4A protobinary system thanks to theoretical predictions of chemical abundances through the passage of shocks. 

Given the lack of secured detections of sulfuretted species in ices, the {pre-shock} abundances of sulfuretted species in interstellar ices are highly uncertain. 
\citet{Palumbo1997} searched for the absorption feature at 4.90 $\mu$m attributed to OCS towards a sample of massive protostars, resulting in  a detection towards three sources with associated solid OCS abundances of $2.3 - 7.0 \times 10^{-8}$ relative to the number of H nuclei. However, upper limits of the OCS abundance down to $3.8 \times 10^{-9}$ have also been derived. 
\citet{Smith1991} searched for the absorption band at 3.90 $\mu$m attributed to H$_2$S towards six protostars, resulting in non-detections only. The derived upper limits of solid H$_2$S abundances between $< 4 \times 10^{-8}$ and $< 1.6 \times 10^{-7}$ are slightly lower but not inconsistent with the atomic sulfur abundance of $1 - 4 \times 10^{-7}$ estimated by \citet{Anderson2013} towards IRAS4A. 
According to \citet{Boogert2015}, solid SO$_2$ has been possibly identified towards only one line of sight so far through a relatively good fit of the absorption band observed at 7.58 $\mu$m with a SO$_2$:CH$_3$OH mixture resulting in an absolute abundance of $\sim 4 \times 10^{-7}$ \citep{Boogert1997}. 

Here, we decide to adopt a two-stage approach. In the first stage, the cold gas-grain chemistry is followed for constant dark cloud physical conditions with the \texttt{MOMICE} gas-grain astrochemical model modified and updated from the \texttt{GRAINOBLE} model \citep{Taquet2012, Taquet2013, Taquet2014, Taquet2016, Dulieu2019} in order to estimate the pre-shock abundances. In the second stage, the physical-chemical evolution impacted by stationary one-dimensional Continuous (C) magnetised shocks is followed using the 2015 version of the Paris-Durham shock model \citep{Flower2015}. 
The dynamical age of the IRAS4A outflows is likely of a few $10^4$ yr, which should be much longer than the equilibrium time of 1000-2000 yr for the shock waves considered in this work, justifying the choice of stationary C-type shocks.  

The gas phase chemical network used in \texttt{MOMICE} for the first stage is based on the 2013 version of the KIDA database updated for reactions involving sulfuretted species following \citet{Smith2004}, \citet{Loison2012}, and \citet{Vidal2018}. The surface chemical network is based on that of \citet{Taquet2016} and is updated following recent works focusing on sulfur chemistry in ices \citep{Vidal2018, Laas2019}. 
The initial elemental abundances are taken from the "low-metal abundances" set EA1 of \citet{Wakelam2008} whilst we set the initial abundance of S$^+$ to $4 \times 10^{-7}$, corresponding to the maximum atomic sulfur abundance derived by \citet{Anderson2013} from S[I] observations towards IRAS4A with {\it Spitzer}. The initial abundances are the following: 
$X$(H$_2$) = 0.5; $X$(He) = $9.00 \times 10^{-2}$; $X$(C$^+$) = $7.3 \times 10^{-5}$; $X$(O) = $1.76 \times 10^{-4}$; $X$(N) = $2.14 \times 10^{-5}$; $X$(Si) = $8.0 \times 10^{-9}$; $X$(Na) = $2.0 \times 10^{-9}$; $X$(Mg) = $7.0 \times 10^{-9}$; $X$(S$^+$) = $4.2 \times 10^{-7}$; $X$(Fe$^+$) = $3.00 \times 10^{-9}$. 
The chemistry is followed for typical molecular cloud physical conditions: a constant density $n_{\rm H} = 2 \times 10^4$ cm$^{-3}$, a visual extinction $A_{\rm V} = 20$ mag, a cosmic ray ionisation rate $\zeta = 1.0 \times 10^{-17}$ s$^{-1}$, and for two constant temperatures $T_{\rm gas} = T_{\rm dust} = 10$ and 20 K until the time reaches $10^6$ yr (i.e. $\sim 3$ times the free-fall time at $n_{\rm H} = 2 \times 10^4$ cm$^{-3}$). 

\begin{figure}[htp]
\centering 
\includegraphics[width=\columnwidth]{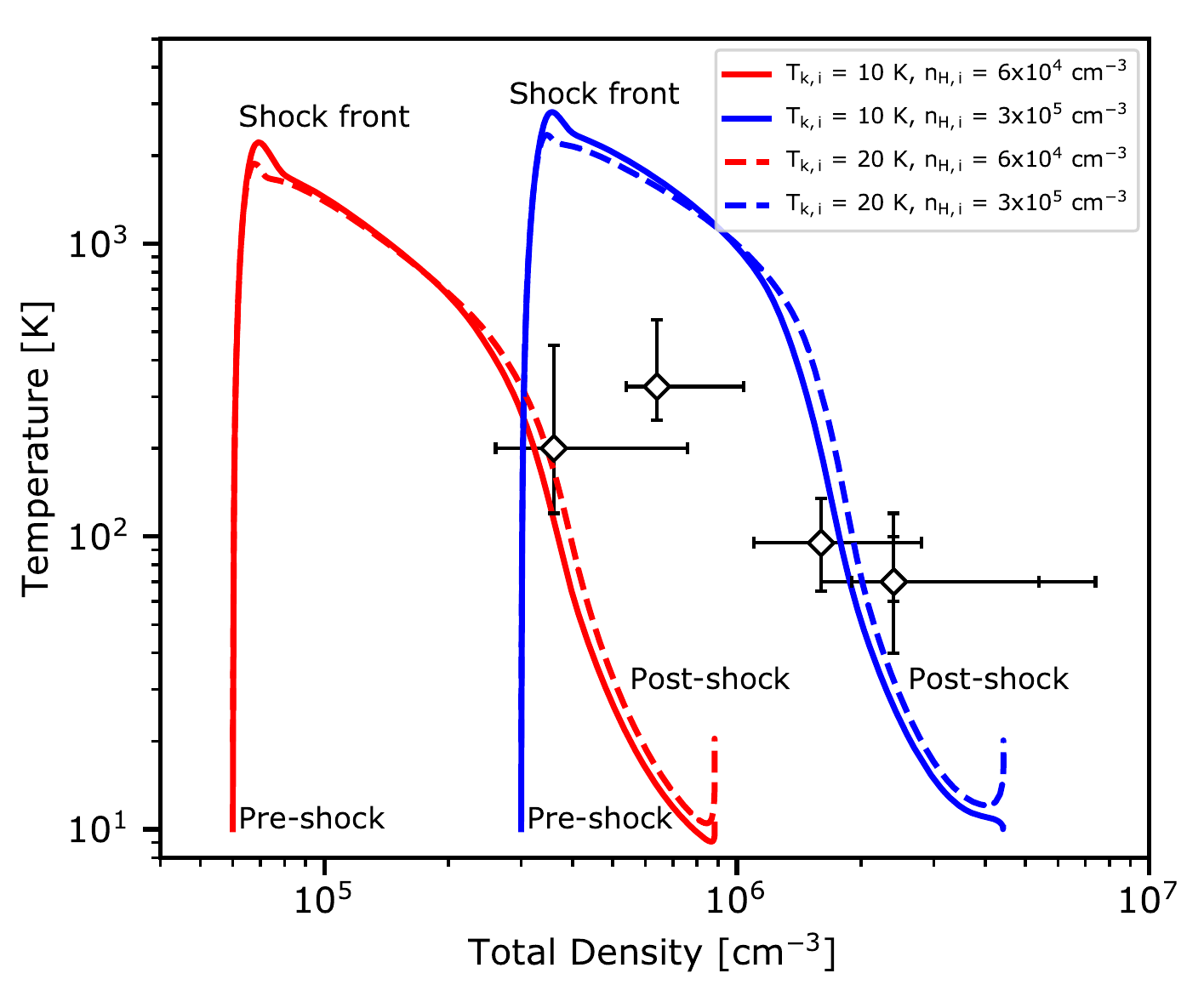}
\caption{
Comparison between the density and temperature evolutions as predicted by the Paris-Durham shock model (coloured curves) for initial {pre-shock} densities of $6 \times 10^4$ and $3 \times 10^5$ cm$^{-3}$ and initial temperatures of 10 and 20 K with the densities and temperatures estimated with the RADEX analysis of the SO$_2$ emission towards various SiO peaks (diamonds; see Table \ref{LVGresults}).
} 
\label{physcond_shock}
\end{figure}

In the second stage, we consider C-shocks with a shock velocity of 20 km s$^{-1}$ and pre-shock (dust and gas) temperatures of 10 and 20 K. To choose the pre-shock densities, we first compared the evolution of density and gas temperature as predicted by the Paris-Durham shock model for various {pre-shock} densities $n_{\rm H,ini}$ with those found with the RADEX analysis of the SO$_2$ emission presented in Section \ref{radex}. As shown in Figure \ref{physcond_shock}, it is found that shocks with $n_{\rm H,ini}$ between $6 \times 10^4$ and $3 \times 10^5$ cm$^{-3}$ reproduce the different observed density and temperature values in the post-shock zones where the gas cools down through radiative cooling. We therefore used these two pre-shock density values for the modelling.
%
%
The chemical network used in the shock code is based on that of \citet{Flower2015} updated for sulfur chemistry following \citet{Smith2004}, \citet{Loison2012}, and \citet{Vidal2018}. In total, the network includes 144 gaseous, icy, or refractory species linked by 1070 chemical processes, and contains only the following sulfuretted species: S, HS, H$_2$S, CS, SO, SO$_2$, OCS, S$^+$, HS$^+$, H$_2$S$^+$, H$_3$S$^+$, CS$^+$, HCS$^+$, SO$^+$, HSO$^+$, HSO$_2^+$, and HOCS$^+$. This network is consequently smaller than the one used in the first stage, but includes the most important formation and destruction reactions for H$_2$S, OCS, CS, SO, and SO$_2$ and for the oxygenated species O, OH, O$_2$, and H$_2$O, that are key for the sulfur chemistry. 

\subsection{Results}

\begin{figure*}[htp]
\centering 
\includegraphics[width=\textwidth]{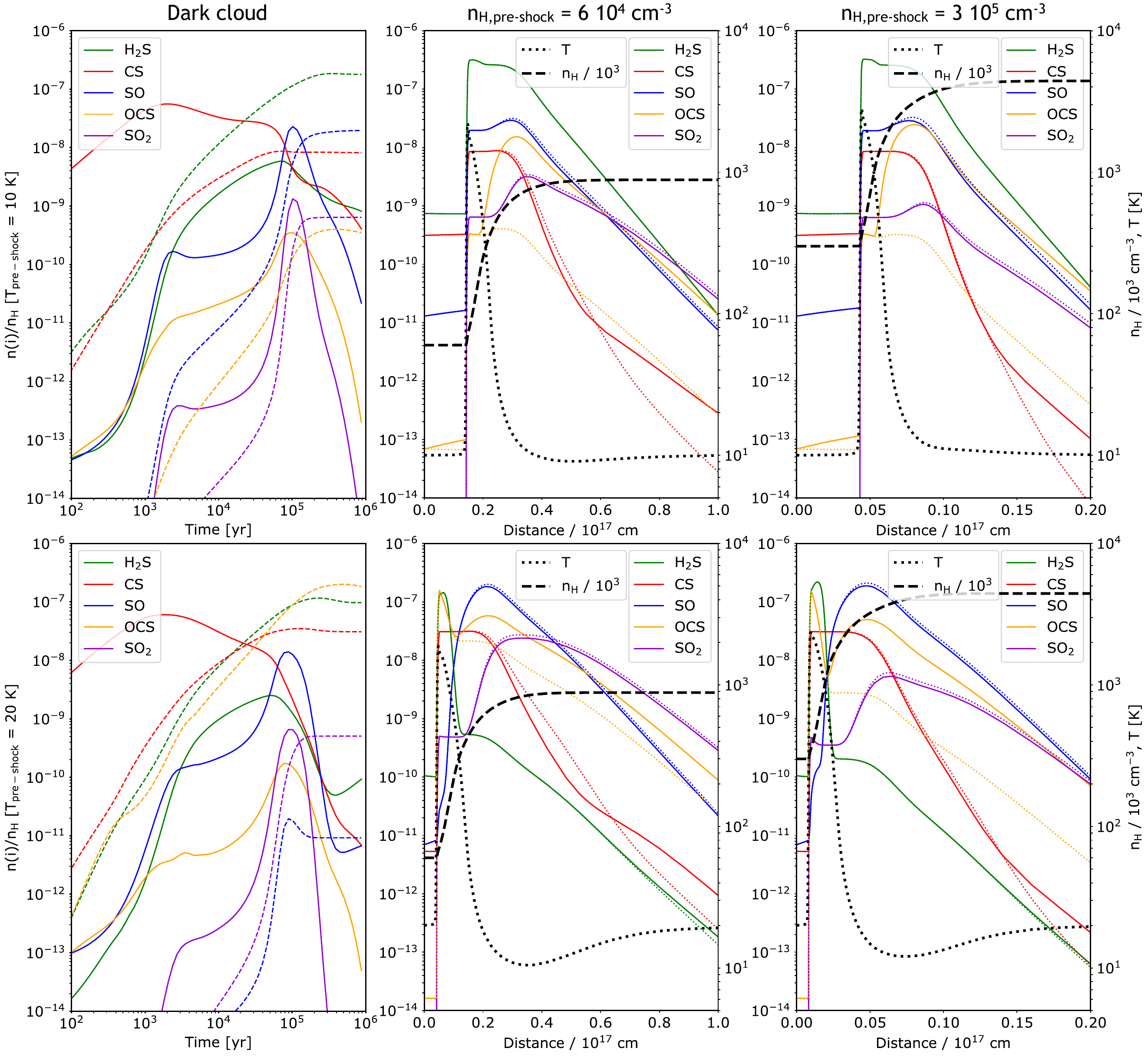}
\caption{
Left panels: Absolute abundances of selected sulfuretted species as a function of time predicted by the \texttt{GRAINOBLE} astrochemical model in the gas phase (solid curves) and in interstellar ices (dashed curves) at 10 K (top) and 20 K (bottom) and for $n_{\rm H} = 2 \times 10^4$ cm$^{-3}$, $\zeta = 10^{-17}$ s$^{-1}$, and $A_{\rm V} = 20$ mag. 
Middle and right panels: Gaseous absolute abundances of selected sulfuretted species (coloured curves) and physical conditions (black curves) profiles in 1D C-type shocks predicted by the Paris-Durham shock code for a pre-shock density of $6 \times 10^4$ cm$^{-3}$ (centre panels) and $3 \times 10^5$ cm$^{-3}$ (right panels). Solid and dashed coloured curves show the abundances with and without the OCS formation/destruction reactions introduced by \citet{Loison2012}, respectively (see text for more details). 
For the shocks considered here, the entire icy material formed in dark clouds is sputtered in the shock front, final ice abundances in the left panels are therefore similar to the gas phase obtained at the shock front in the middle and right panels, with the exception of H$_2$S that is efficiently formed through hot gas phase chemistry. 
} 
\label{results_grainoble}
\end{figure*}

The left panels of Figure \ref{results_grainoble} show the temporal evolution of the absolute abundances of the main sulfuretted species in ices and in the gas phase predicted by the \texttt{MOMICE} astrochemical model at n$_{\rm H}$ = $2 \times 10^4$ cm$^{-3}$, $T = 10$ and 20 K, $\zeta = 10^{-17}$ s$^{-1}$, and $A_{\rm V} = 20$ mag. 
H$_2$S is the one of the most abundant sulfuretted molecules in ices with a final abundance of $\sim 2 \times 10^{-7}$ and $\sim 1 \times 10^{-7}$ at 10 and 20 K, respectively. These values are similar to the observed upper limits in abundances derived by \citet{Smith1991}. The surface chemistry of H$_2$S is thought to be similar to that of water, atomic sulfur is efficiently hydrogenated through addition reactions involving H atoms through HS, assumed to be barrierless. 
The abundance of solid SO dramatically decreases from $10^{-8}$ to $2 \times 10^{-11}$ between 10 and 20 K because of the increasing diffusion of carbon atoms along the surface that increases the reactivity of the SO+C reaction forming CO+S.
On the other hand, the abundance of solid OCS increases strongly with the temperature, from $4 \times 10^{-10}$ to $2 \times 10^{-7}$, between 10 and 20 K. Like CO$_2$, solid OCS formation is triggered for moderate temperatures of 15-20 K since the relatively heavy species S, CO and CS start to diffuse along the surface and react with radicals such as OH and HS to form OCS. 
CS is the third most abundant molecule in ices with an abundance of $1-3 \times 10^{-8}$. Unlike H$_2$S, most solid CS has a gas phase origin and is then efficiently trapped "intact" in ices after its freeze-out. Indeed, its most important destruction reaction on ices, converting solid CS to solid HCS, has a limited efficiency because it is thought to have an activation barrier of $\sim 1000$ K \citep[see][]{Vidal2018}. 
The abundance of SO$_2$ is about $10^{-9}$ between 10 and 20 K and is mostly due to gas phase chemistry followed by freeze-out. The predicted abundance is two orders of magnitude lower than the solid SO$_2$ abundance derived by \citet{Boogert1997} towards the massive protostar W33A. However, according to \citet{Boogert2015} SO$_2$ has only possibly been identified so far, and more secured detections are needed to confirm the presence of SO$_2$ in ices.

The final abundances predicted by \texttt{MOMICE} are considered as initial pre-shock abundances in the shock code.  
The centre and right panels of Fig. \ref{results_grainoble} show the spatial evolution of the gas phase abundances of the sulfuretted species within shocks for pre-shock densities $n_{\rm H,ini}$ of $6 \times 10^4$ and $3 \times 10^5$ cm$^{-3}$, pre-shock temperatures of 10 and 20 K, and a shock velocity of 20 km s$^{-1}$. 
The compression of the material at the shock front induces an increase of the total density and a sudden increase of the gas temperature between 2000 and 3000 K, the exact value depending on the shock velocity, the initial density, and the pre-shock temperature. The dust temperature is assumed to remain constant over time, and remains equal to the initial temperature of 10 or 20 K. Behind the shock front, radiative cooling efficiently decreases the temperature of the gas down to its initial temperature. The electronic abundance at $T_{\rm pre-shock} = 20$ K predicted by \texttt{MOMICE} is higher than at 10 K by a factor of two, increasing the coupling between the neutral and ionised gas, and inducing a faster compression of the gas.

For the shock velocity of 20 km s$^{-1}$ considered here, the entire icy material is sputtered into the gas phase within the shock front. { However, the refractory material remains intact after the passage of such shocks since refractory grains start to be efficiently destroyed through sputtering for shocks faster than typical $30-40$ km s$^{-1}$ \citep[see][]{May2000}. }
Once sputtered, H$_2$S and OCS can be efficiently destroyed in the gas through the H$_2$S+H and OCS+H abstraction reactions since they have moderate activation barriers of 860 and 2000 K, respectively \citep{Schofield1973, Tsunashima1975}. Such barriers can be overcome at temperatures higher than 1000 K reached in the shock front for about 30-80 yr, depending on the pre-shock density. An initial temperature of 20 K increases the pre-shock abundance of atomic H in the gas phase by about two orders of magnitude because H atoms cannot deplete efficiently on grains, increasing the efficiency of these destruction reactions of H$_2$S and OCS. 
Reactions between atomic H and SO or SO$_2$, on the other hand, have activation barriers higher than 10000 K \citep{Leen1988, Blitz2006}. SO and SO$_2$ can therefore survive during their passage in the shock front for the shocks considered here. 

Behind the shock front, the abundance of SO increases with the distance from the shock, especially for the models considering an initial temperature of 20 K where SO has a low initial abundance. SO$_2$ is then formed later on at larger distances with respect to the shock front. 
SO is mainly formed through three barrierless neutral-neutral S+OH, HS+O, and S+O$_2$ reactions in the warm gas. All species involved in SO formation have moderately high abundances, between $10^{-9}$ and $5 \times 10^{-7}$, after the shock front, and allow for an efficient formation of SO. 
SO is then gradually converted into SO$_2$ mainly through the SO+OH barrierless reaction. SO$_2$ reaches its maximal abundance behind the shock at a distance of $3-4 \times 10^{16}$ and $6-8 \times 10^{15}$ cm for pre-shock densities of $6 \times 10^4$ and $3 \times 10^5$ cm$^{-3}$, respectively. 
Unlike SO and SO$_2$, H$_2$S cannot be efficiently formed in the gas phase behind the shock and its abundance gradually decreases due to freeze-out onto dust grains.

For OCS, the situation is more complex. \citet{Loison2012} recently introduced a few reactions that are potentially important for OCS formation and destruction, notably the CS+OH, SO+CH, S+HCO, OCS+CH, and OCS+C reactions. 
The CS+OH was already present in astrochemical networks but with a low rate \citep[$\alpha = 9.4 \times 10^{-14}$ cm$^{-3}$ s$^{-1}$; $\gamma = 800$ K;][]{Leen1988} whilst the SO+CH and S+HCO reactions were not present in astrochemical networks, \citet{Loison2012} estimated them to be reactive with rates of $\sim 1 \times 10^{-10}$ cm$^{-3}$ s$^{-1}$. 
Dotted coloured lines in the centre and right panels of Fig. \ref{results_grainoble} show the chemical abundances without the new reactions introduced by \citet{Loison2012}. It can be seen that, without these new reactions, OCS is not efficiently formed in the post-shock gas phase and OCS would be mostly present in the shock front, similar to H$_2$S { \citep[as recently suggested  by][]{Podio2014, Holdship2016}.}, especially at $T_{\rm pre-shock} = 20$ K when the OCS abundance is high. On the other hand, the incorporation of the reactions suggested by \citet{Loison2012} triggers the formation of gaseous OCS behind the shock and OCS behaves like SO. For a pre-shock temperature of 20 K, the OCS abundance first drops because of "hot" gas-phase chemistry before re-increasing due to the barrierless CS+OH and SO+CH reactions. 
After its sputtering in the shock front, CS is not significantly affected by hot nor warm gas phase chemistry, its abundance remains constant near the shock before decreasing due to freeze-out onto dust grains.

The gradual transition between H$_2$S, OCS, SO, and SO$_2$ shown here confirms the earlier predictions by \citet{PineaudesForets1993} in shocks and by \citet{Charnley1997} and \citet{Wakelam2004} in hot cores. 
The main differences with respect to the work \citet{PineaudesForets1993} is the incorporation of OCS in the chemical network and the H$_2$S formation in cold interstellar ices subsequently sputtered in the shock front in our work whilst \citet{PineaudesForets1993} formed H$_2$S in the shock front from ion-neutral reactions through SH$^+$ and H$_2$S$^+$.

In conclusion, the modelling presented here suggests that H$_2$S should trace the shock front where the icy material has been recently sputtered whilst SO$_2$ would rather trace the post-shock regions. This would explain why SO$_2$ is mostly detected towards the NE-SW outflow driven by IRAS4A2 since this outflow is much more elongated and consequently likely older. SO and CS should be abundant both in the shock front and behind the shock regardless of the shock velocities, explaining their bright and extended emission throughout the two outflows. However, the evolution of the SO/CS column density ratio along the outflows, especially near SiO peak 2, could be explained by a different chemical behaviour behind shocks, SO being formed at moderate gas phase temperatures unlike CS. 

With the new reactions introduced by \citet{Loison2012}, the OCS abundance profiles along the shocks are complex and mostly depend on the initial pre-shock temperature that governs the OCS abundance in ices. Our observations show that OCS is much brighter in the shorter, and likely younger, south outflow driven by IRAS4A1 similarly to CH$_3$OH as seen in Fig. \ref{moment0_all}. These observations would agree well with the models considering a pre-shock temperature of $\sim 20$ K, where OCS is mostly abundant in shock fronts and { the OCS maps do not show any evidence for efficient gas phase formation behind shocks through the reactions suggested by \citet{Loison2012} since we do not observe any enhanced emission on larger scales, like SO for instance.} 
%
%

The conclusions given by the present modelling strongly rely on the predicted abundances of sulfuretted species locked in interstellar ices, which are still poorly constrained by observations. New infrared observations of interstellar ices will soon be carried out with the James Webb Space Telescope at high sensitivity and spectral resolution, likely resulting in more robust estimates of the sulfur reservoir in ices. 
Observing H$_2$S towards the IRAS4A system at a similar angular resolution would also help us to disentangle the puzzling chemistry of OCS in outflows.


\section{Conclusions}

This work presents multi-line interferometric observations of various sulfuretted species towards the outflows driven by the NGC1333-IRAS4A protobinary system with the IRAM-NOEMA interferometer. We analysed the spatial and kinematical evolutions along the outflows of the emission from various OCS, CS, SO, and SO$_2$ transitions detected at different frequencies in the 1.3, 2, and 3 mm bands for two typical angular resolutions of 1 and 2-4$\arcsec$. We used clustering methods and radiative models to evaluate the evolution of the physical and chemical conditions. We also used the state-of-the-art one-dimensional Paris-Durham shock model to study the sulfur chemistry in shocks with the goal of interpreting the chemical differentiation observed along the outflows.
We summarise here the main results of our work:

\noindent
1) OCS, CS, SO, SO$_2$ clearly show different spatial emission along the two outflows driven by IRAS4A. 
Integrated emission maps reveal that OCS is brighter in the south outflow driven by IRAS4A1 whereas SO$_2$ is rather detected along the outflow driven by IRAS4A2 extended along the north east - south west (NE-SW) direction. CS and SO show extended emission along the two outflows.

\noindent
2) SO is detected at extremely-high velocity up to $+25$ km s$^{-1}$ relative to the source velocity, clearly allowing us to distinguish the outflow driven by IRAS4A1 along the north - south direction from the NE-SW outflow driven 4A2. Other sulfuretted species are mostly detected at lower velocities. OCS in particular shows an intriguing "blue" low velocity ($< 5$ km s$^{-1}$ relative to source velocity) emission peak at the beginning of the south outflow traced by SiO and SO. 

\noindent
3) Column density ratio maps have been estimated from a rotational diagram analysis and allowed us to confirm a clear gradient of the OCS/SO$_2$ column density ratio between the outflow driven by IRAS4A1 and the one triggered by 4A2. 

\noindent
4) Non-LTE analysis of four SO$_2$ transitions towards several SiO emission peaks suggests that associated gas should be relatively dense, with densities higher than $10^5$ cm$^{-1}$, and relatively warm ($T > 100$ K) temperatures in most cases.  

\noindent
5) A comparison with theoretical predictions carried out with the state-of-the-art Paris-Durham shock model allowed us to give an interpretation of the chemical evolution of sulfur chemistry along the two outflows, in spite of the complex chemistry at work. 
OCS would have been efficiently formed in "warm" interstellar ices, namely at $T \sim 20$ K, and then recently released in the gas phase of shock through sputtering processes, explaining its bright emission in the likely young south outflow driven by IRAS4A1. In contrast, SO$_2$ should be rather formed behind the shock front through warm gas phase chemistry, explaining why SO$_2$ is mostly detected towards the more elongated and likely older NE-SW outflow driven by IRAS4A2. SO and CS should be abundant both in the shock front and behind the shock regardless of the shock conditions, explaining their bright and extended emission throughout the two outflows.


\begin{acknowledgements}

V.T. is grateful to Sylvie Cabrit and Guillaume Pineau des Forêts for stimulating discussions on the chemistry in shocks.
The authors acknowledge the CALYPSO consortium for the use of the CALYPSO dataset.
This work is based on observations carried out with the IRAM PdBI/NOEMA Interferometer under project numbers V05B and V010 (PI: M.V. Persson), U003 (PI: V. Taquet), and L15AA (PI: C. Ceccarelli and P. Caselli). 
IRAM is supported by INSU/CNRS (France), MPG (Germany) and IGN (Spain).
V.T. acknowledges the financial support from the European Union's Horizon 2020 research and innovation programme under the Marie Sklodowska-Curie grant agreement n. 664931.
This work was supported by (i) the PRIN-INAF 2016 "The Cradle of Life - GENESIS-SKA (General Conditions in Early Planetary Systems for the rise of life with SKA)", (ii) the European Research Council (ERC) under the European Union's Horizon 2020 research and innovation programme, for the Project “The Dawn of Organic Chemistry” (DOC), grant agreement No 741002, and (iii) the European MARIE SKŁODOWSKA-CURIE ACTIONS under the European Union's Horizon 2020 research and innovation programme, for the Project “Astro-Chemistry Origins” (ACO), Grant No 811312. 
C.F. acknowledges support from the French National Research Agency in the framework of the Investissements d’Avenir program (ANR-15- IDEX-02), through the funding of the "Origin of Life" project of the Univ. Grenoble-Alpes.

\end{acknowledgements}

 \appendix

\section{Intensity maps}




\begin{figure*}[htp]
\centering 
\includegraphics[width=1.464\columnwidth]{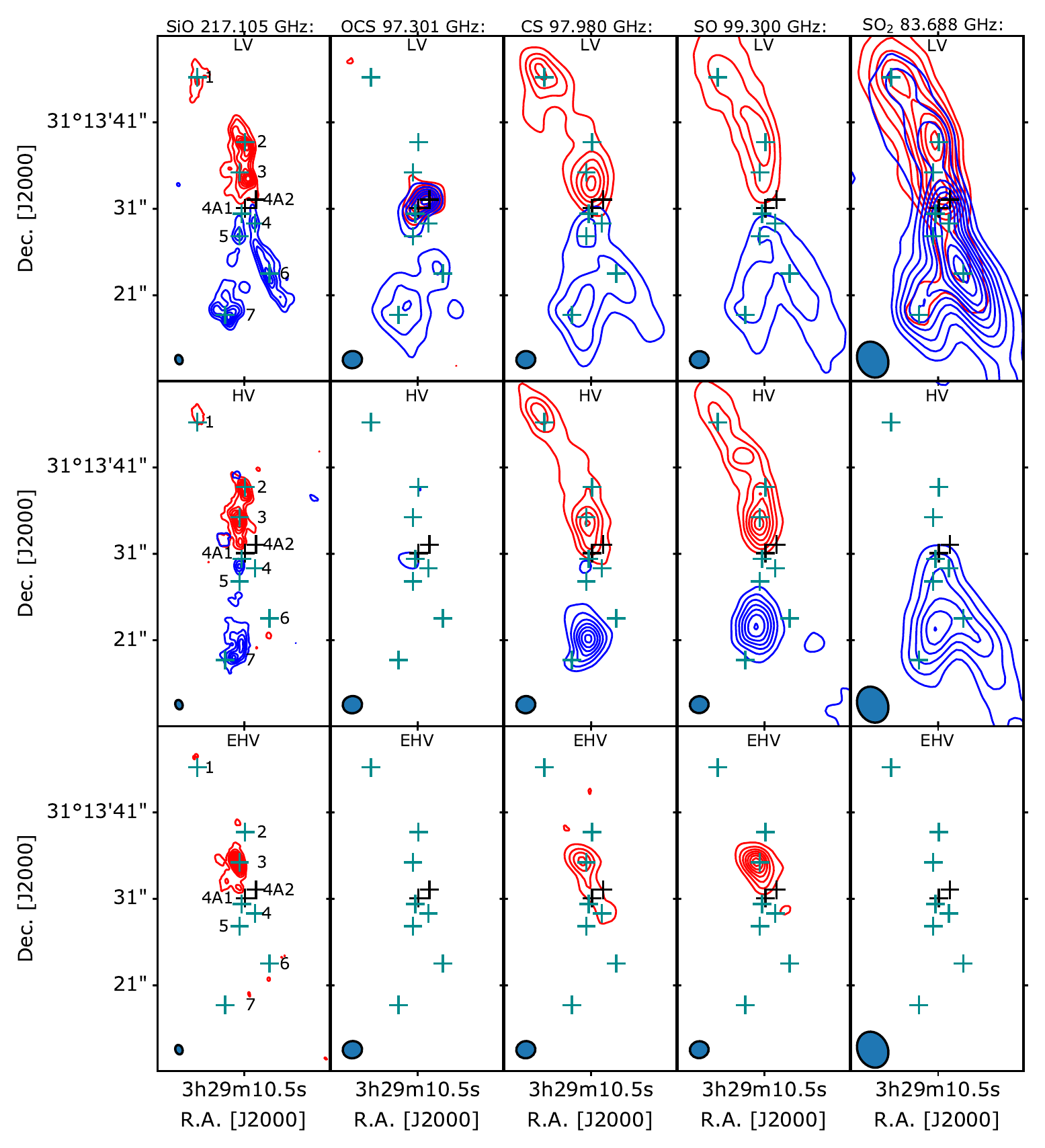}
\caption{Channel maps of the main sulfuretted isotopologues observed at low ($\sim$ 2-4$\arcsec$) resolution for three velocity ranges: low-velocity LV (-10, +5) and (+10, +25) km s$^{-1}$, high-velocity HV (-20, -11) and (+26, +40) km s$^{-1}$, and extremely-high-velocity EHV (-35, -21) and (+41, +55) km s$^{-1}$.} 
\label{channelmaps_lowresol}
\end{figure*}

\begin{figure*}[htp]
\centering 
\includegraphics[width=1.705\columnwidth]{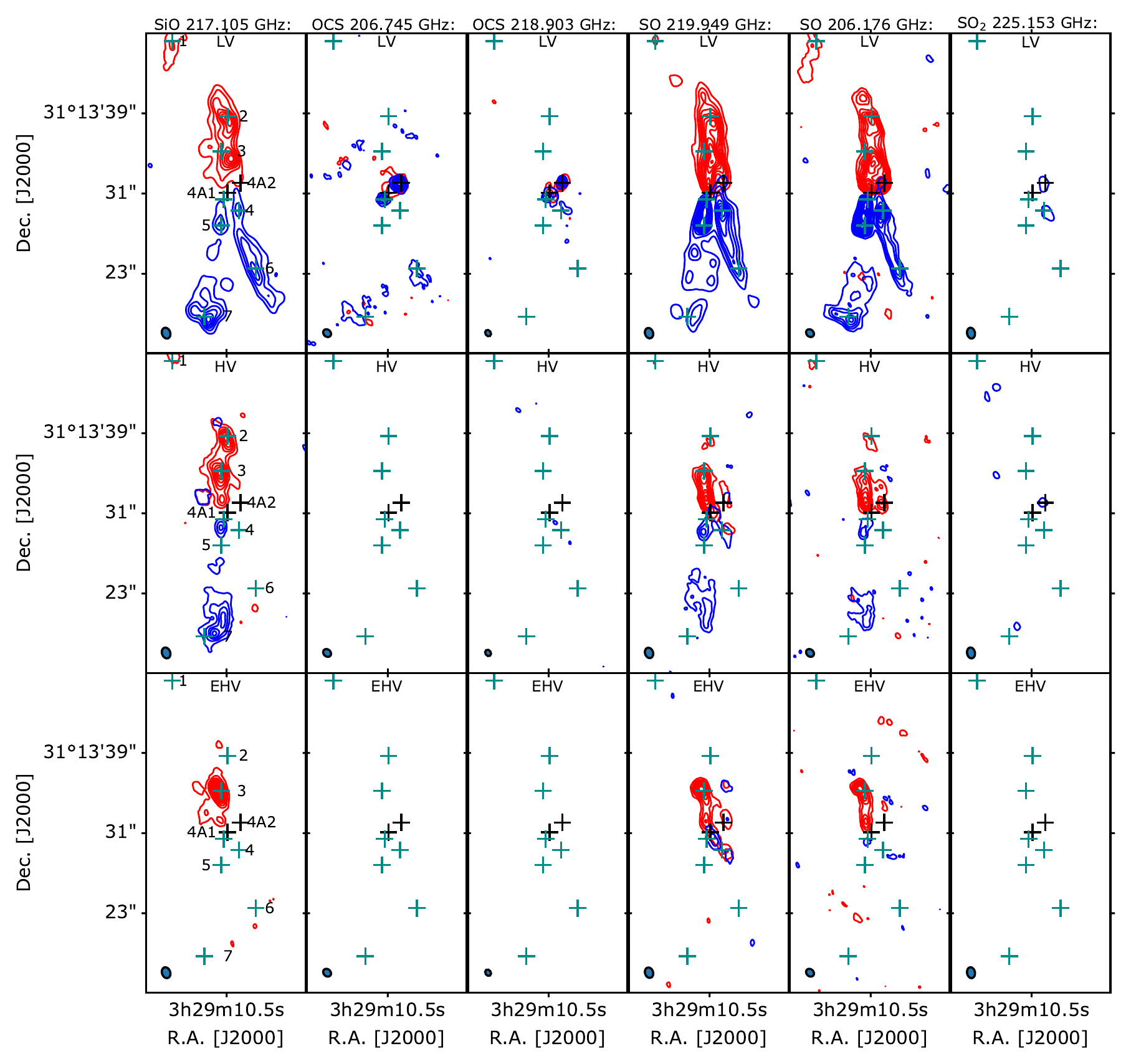}
\caption{Channel maps of the main sulfuretted isotopologues observed at high ($\sim$ 1$\arcsec$) resolution for three velocity ranges: low-velocity LV (-10, +5) and (+10, +25) km s$^{-1}$, high-velocity HV (-20, -11) and (+26, +40) km s$^{-1}$, and extremely-high-velocity EHV (-35, -21) and (+41, +55) km s$^{-1}$.} 
\label{channelmaps_highresol}
\end{figure*}

\section{Spectra towards SiO and OCS peaks}

\begin{figure*}[htp]
\centering 
\includegraphics[width=0.9\columnwidth]{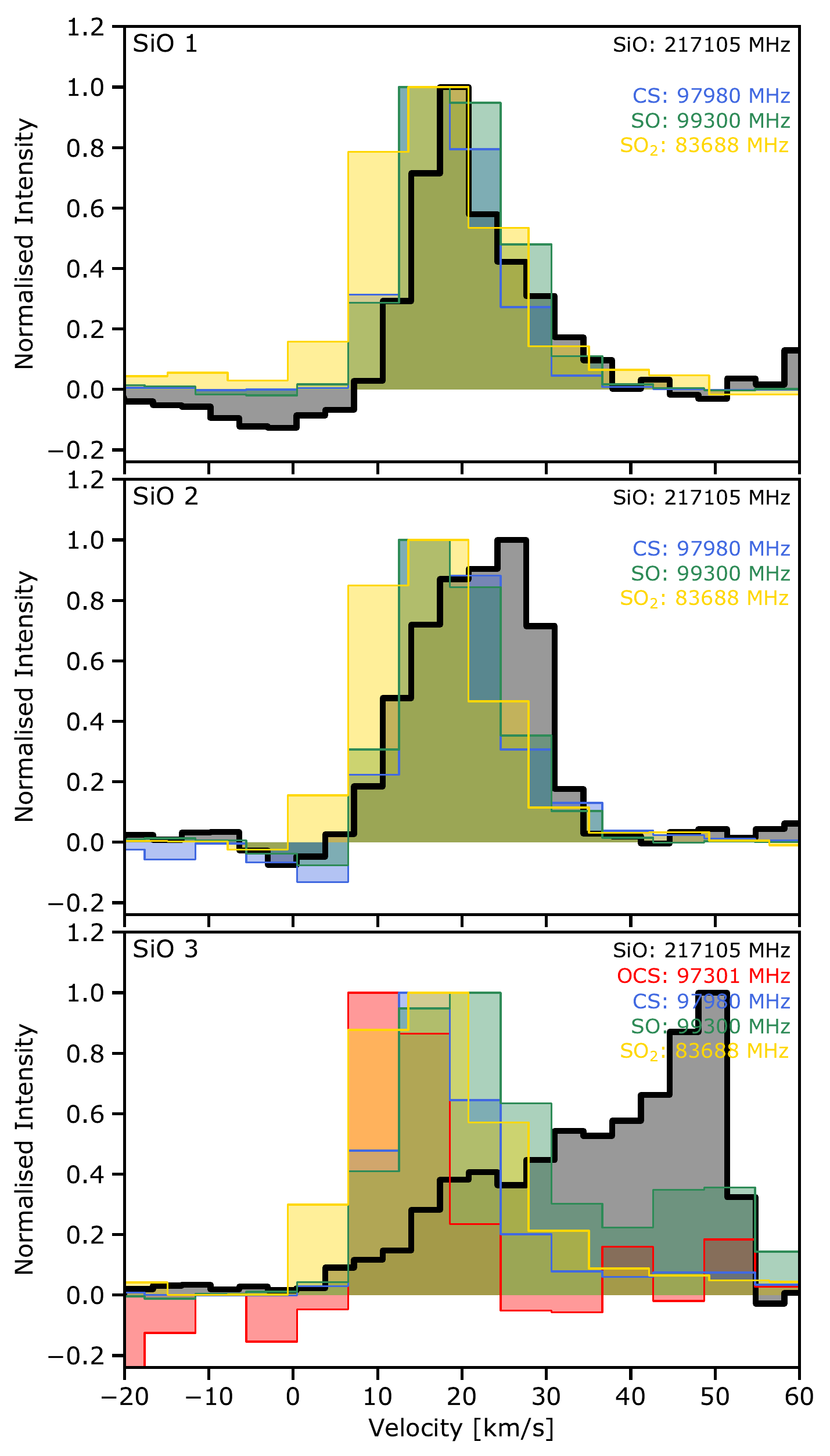}
\includegraphics[width=0.9\columnwidth]{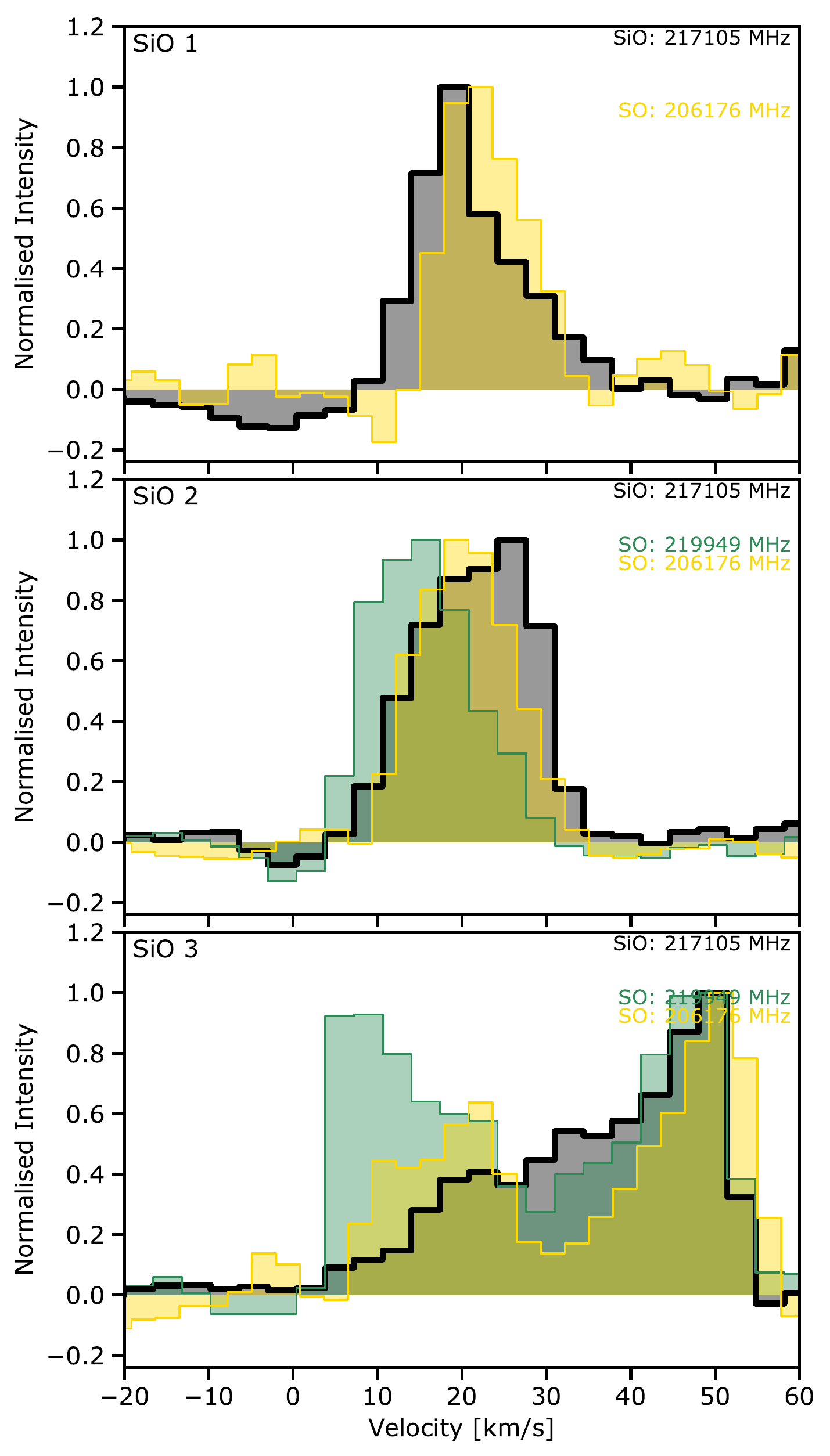}
\caption{Normalised spectra of sulfuretted transitions observed at low (left panels) and high (right panels) angular resolution compared with the normalised SiO spectra at high angular resolution towards the SiO peaks in the north from IRAS4A if the intensity peak is higher than three times the $\sigma$ rms noise level. } 
\label{spectra_sionorth}
\end{figure*}

\begin{figure*}[htp]
\centering 
\includegraphics[width=0.9\columnwidth]{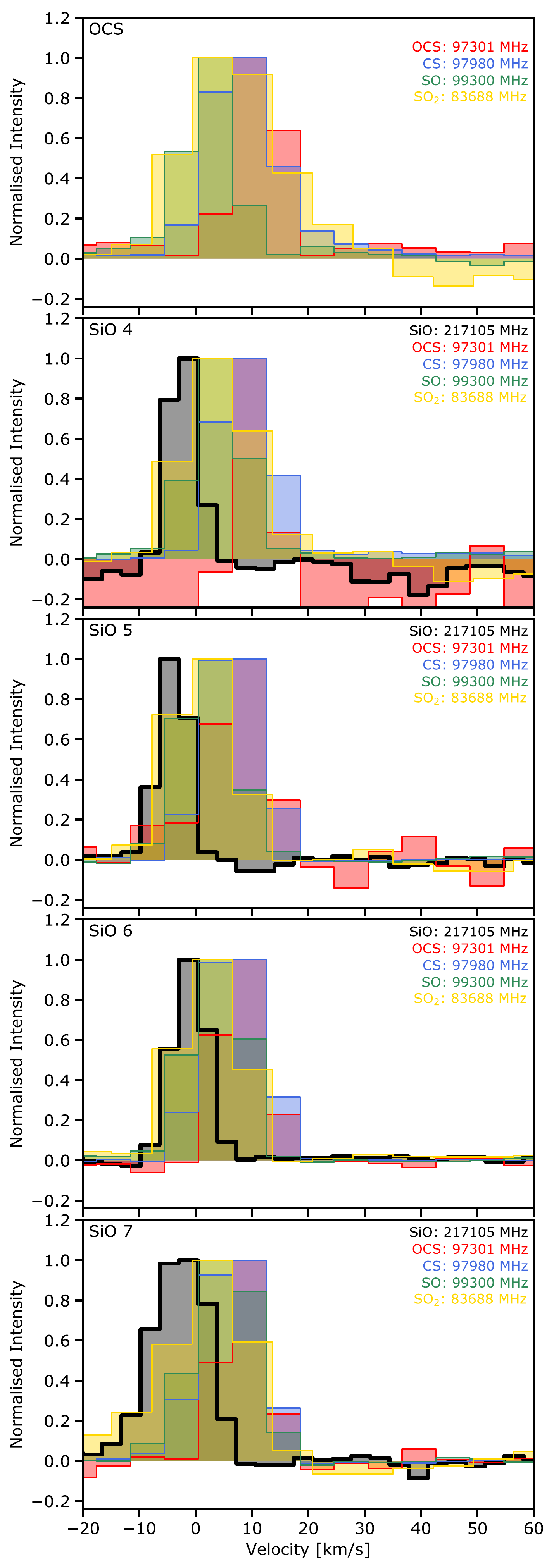}
\includegraphics[width=0.9\columnwidth]{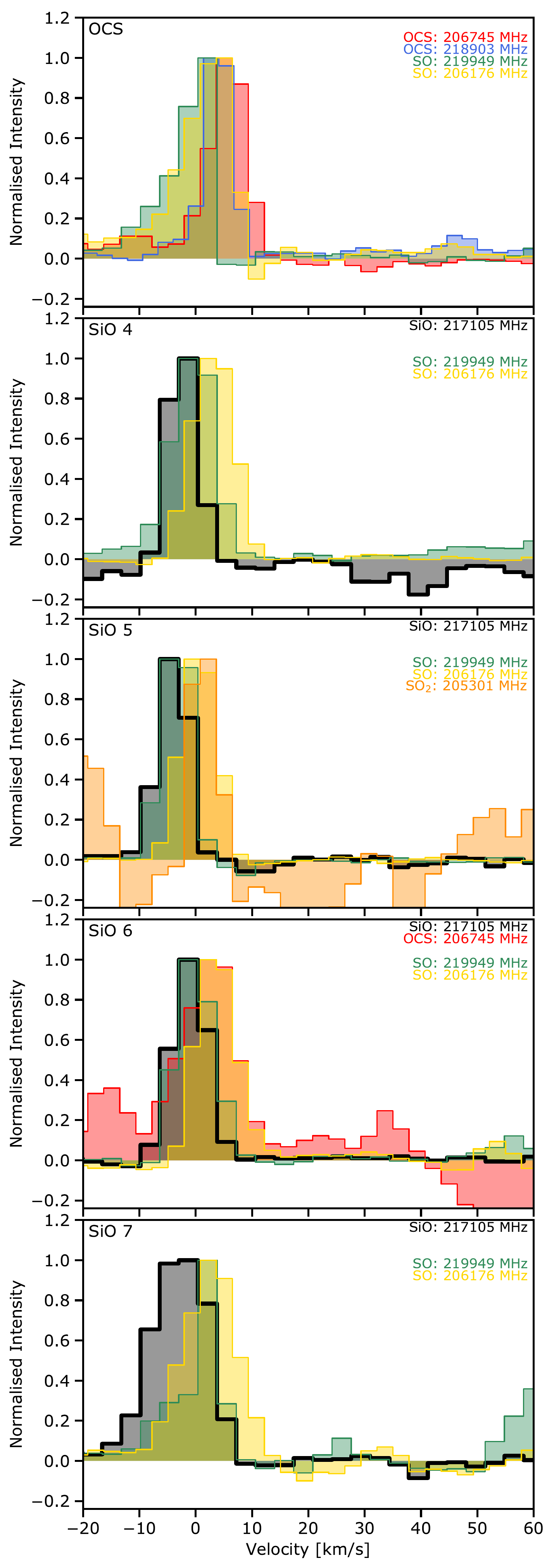}
\caption{Normalised spectra of sulfuretted transitions observed at low (left panels) and high (right panels) angular resolution compared with the normalised SiO spectra at high angular resolution towards the SiO peaks and the OCS peak in the south from IRAS4A if the intensity peak is higher than three times the $\sigma$ rms noise level.} 
\label{spectra_siosouth}
\end{figure*}

\section{Comparison between IRAM-30m and NOEMA spectra}

\begin{figure*}[htp]
\centering 
\includegraphics[width=0.33\textwidth]{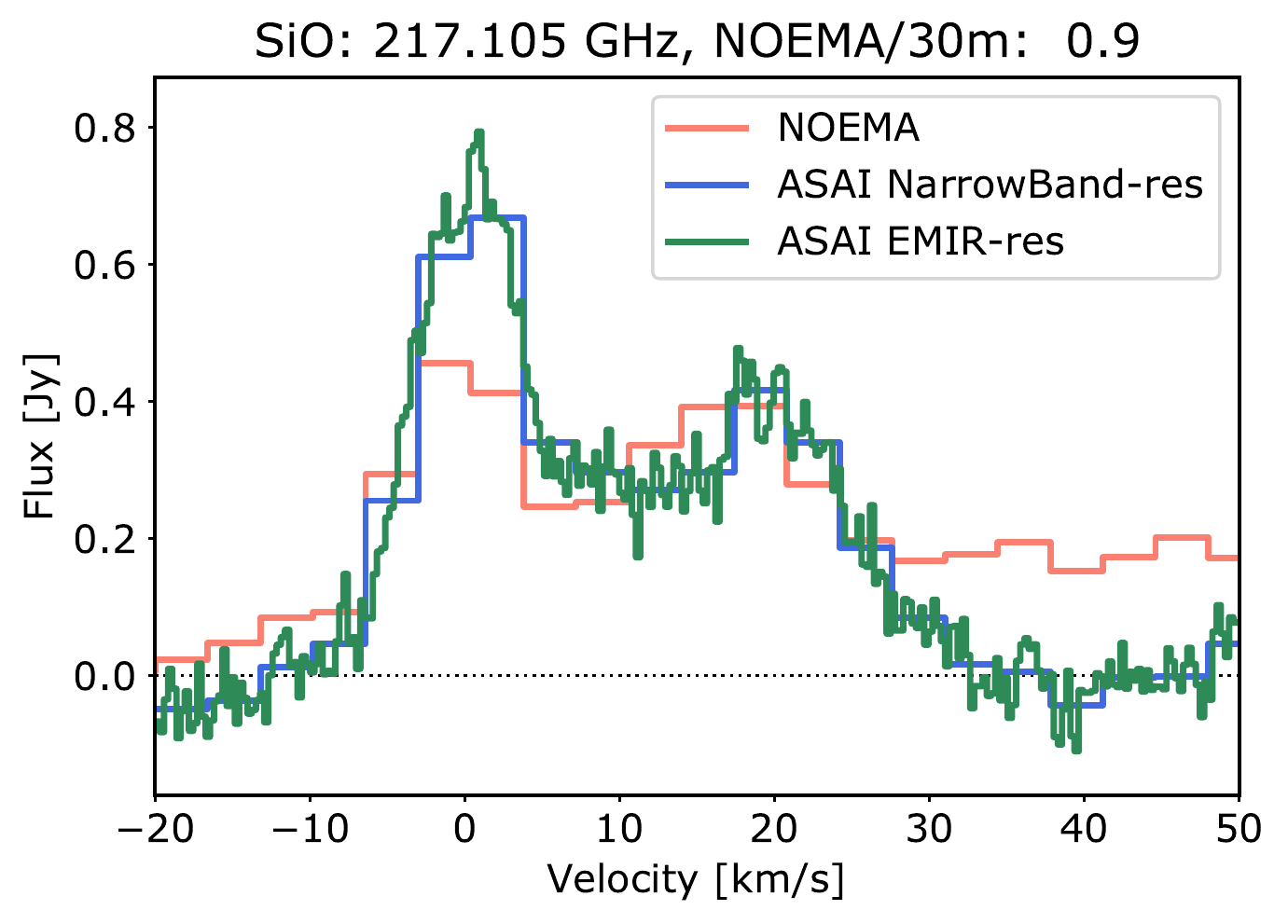}
\includegraphics[width=0.33\textwidth]{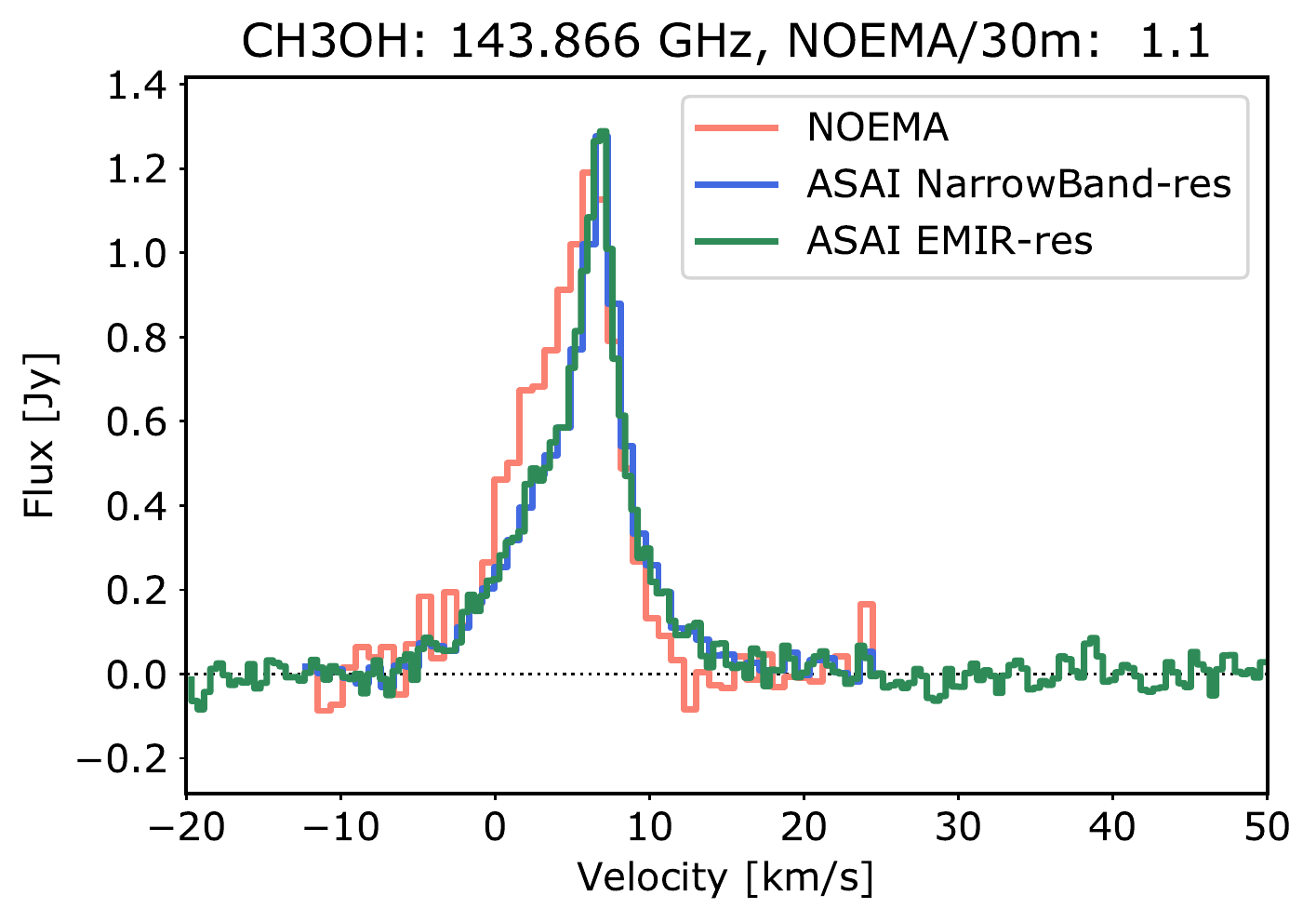} \\
\includegraphics[width=0.33\textwidth]{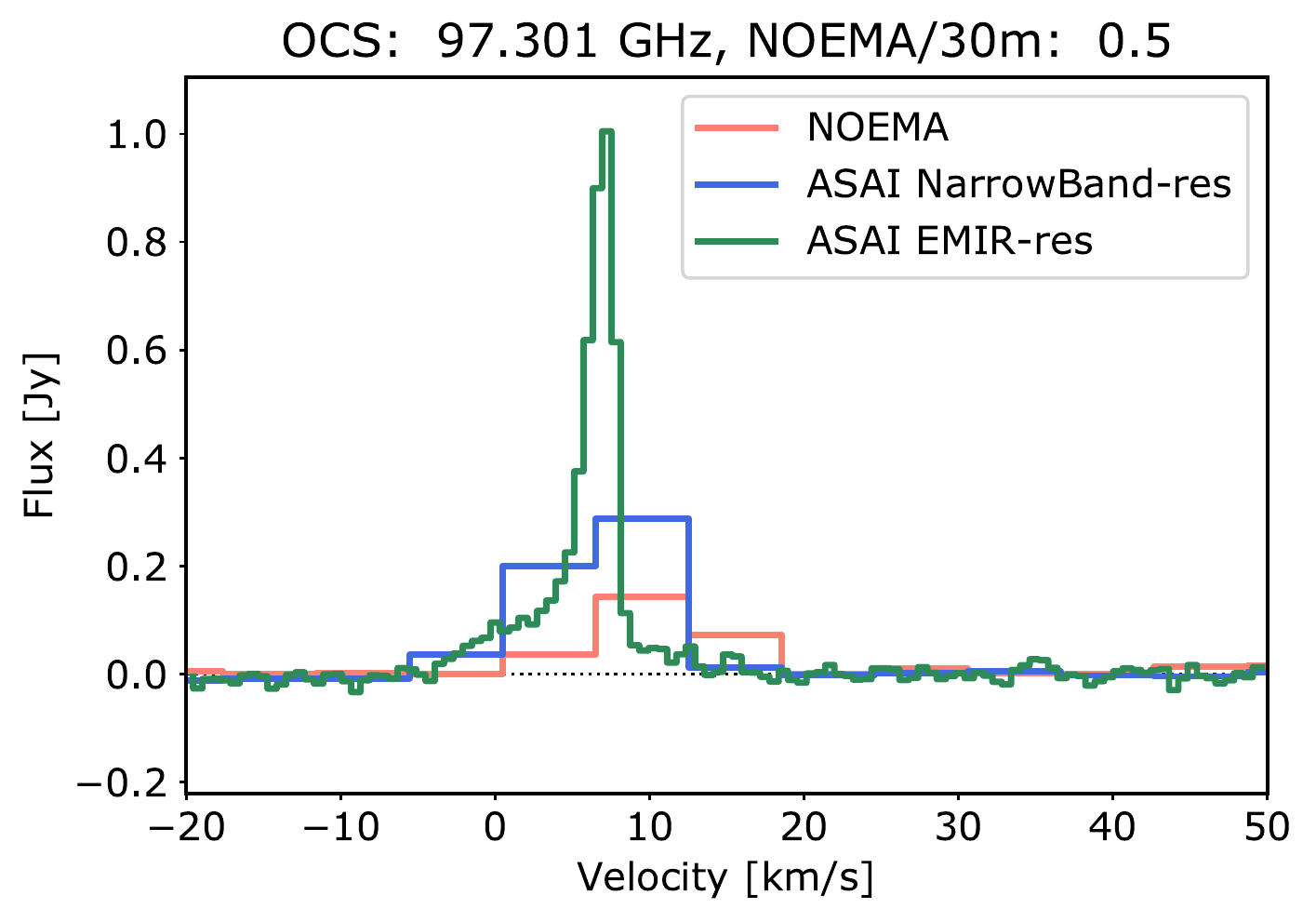} 
\includegraphics[width=0.33\textwidth]{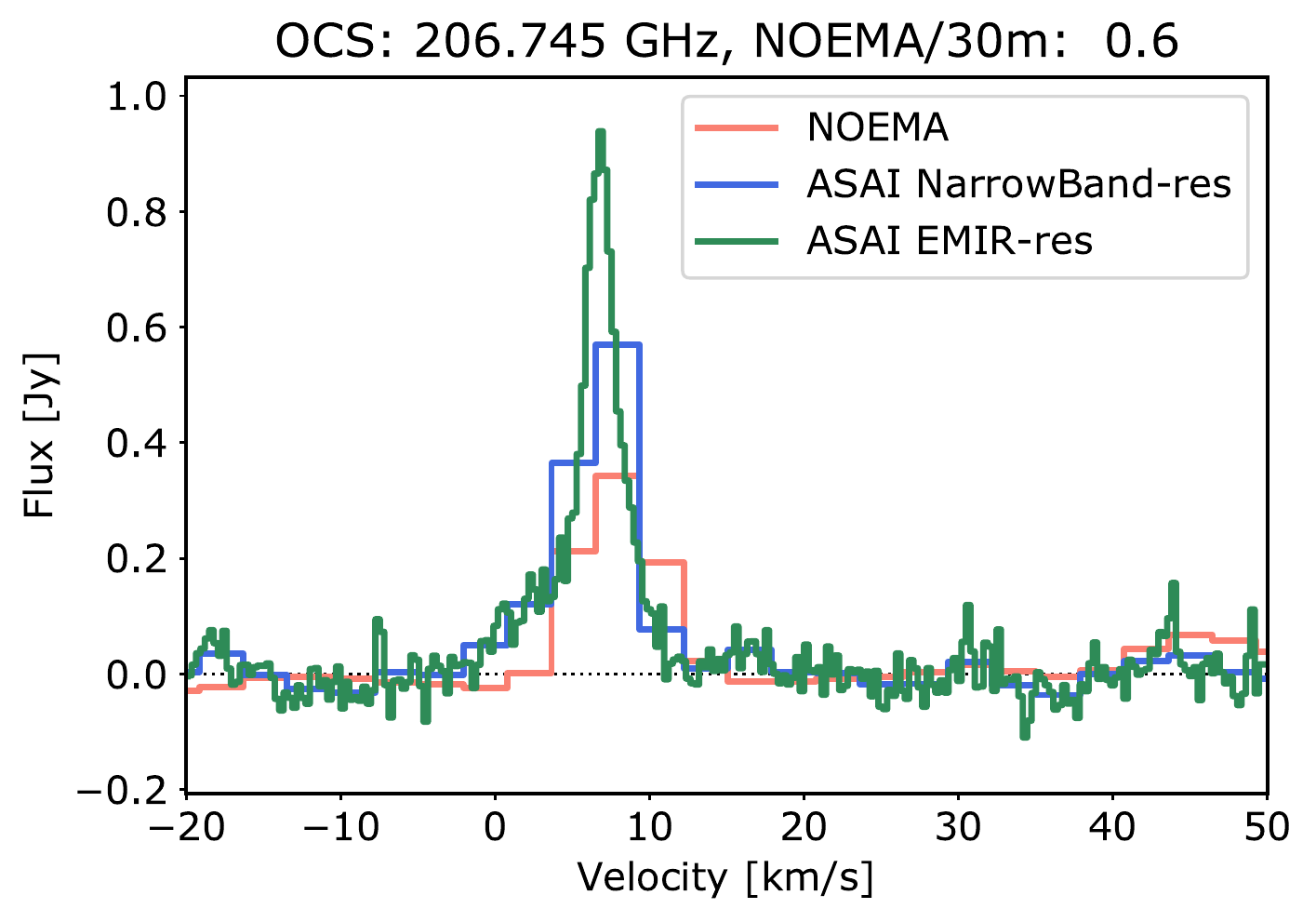}
\includegraphics[width=0.33\textwidth]{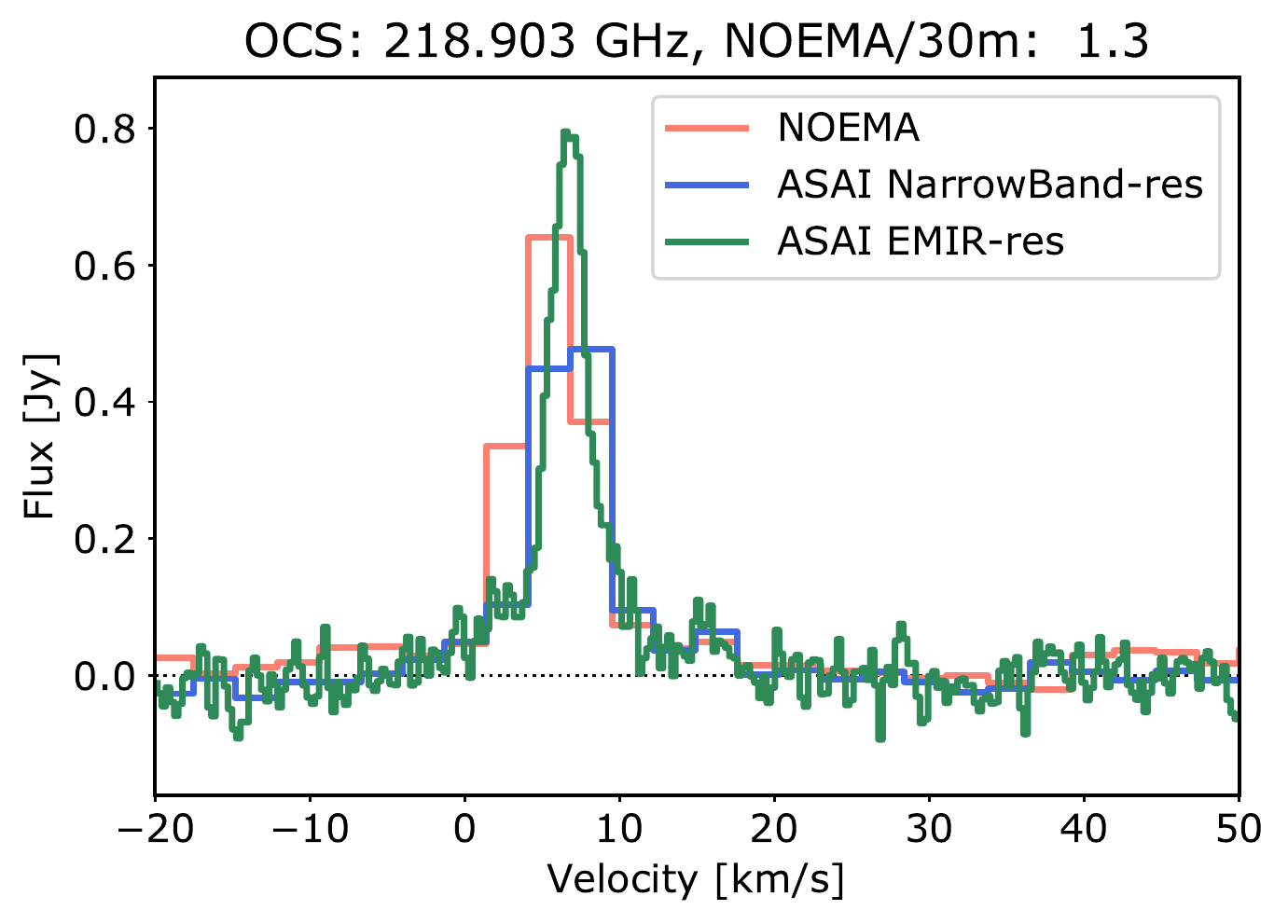} \\
\includegraphics[width=0.33\textwidth]{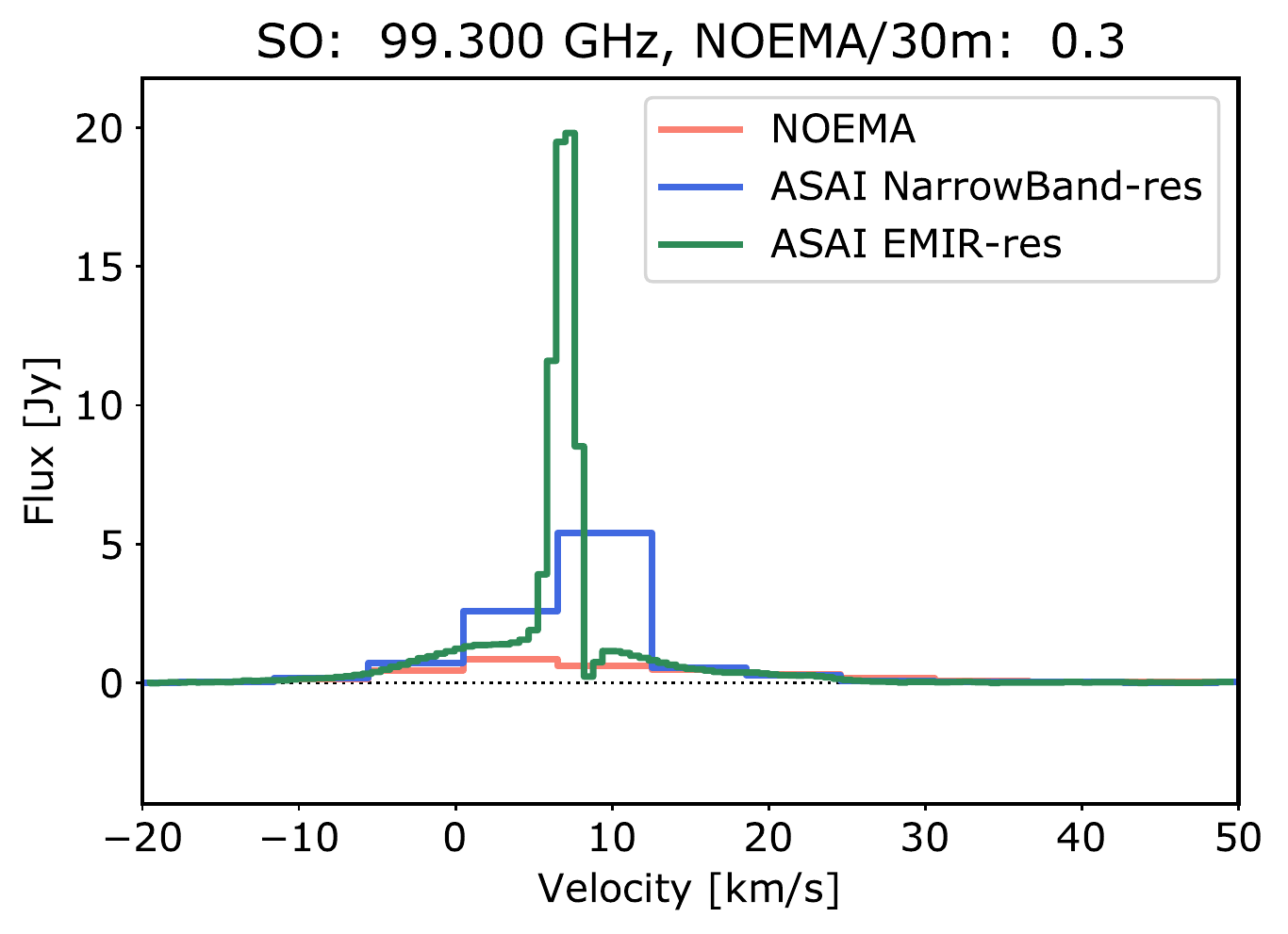}
\includegraphics[width=0.33\textwidth]{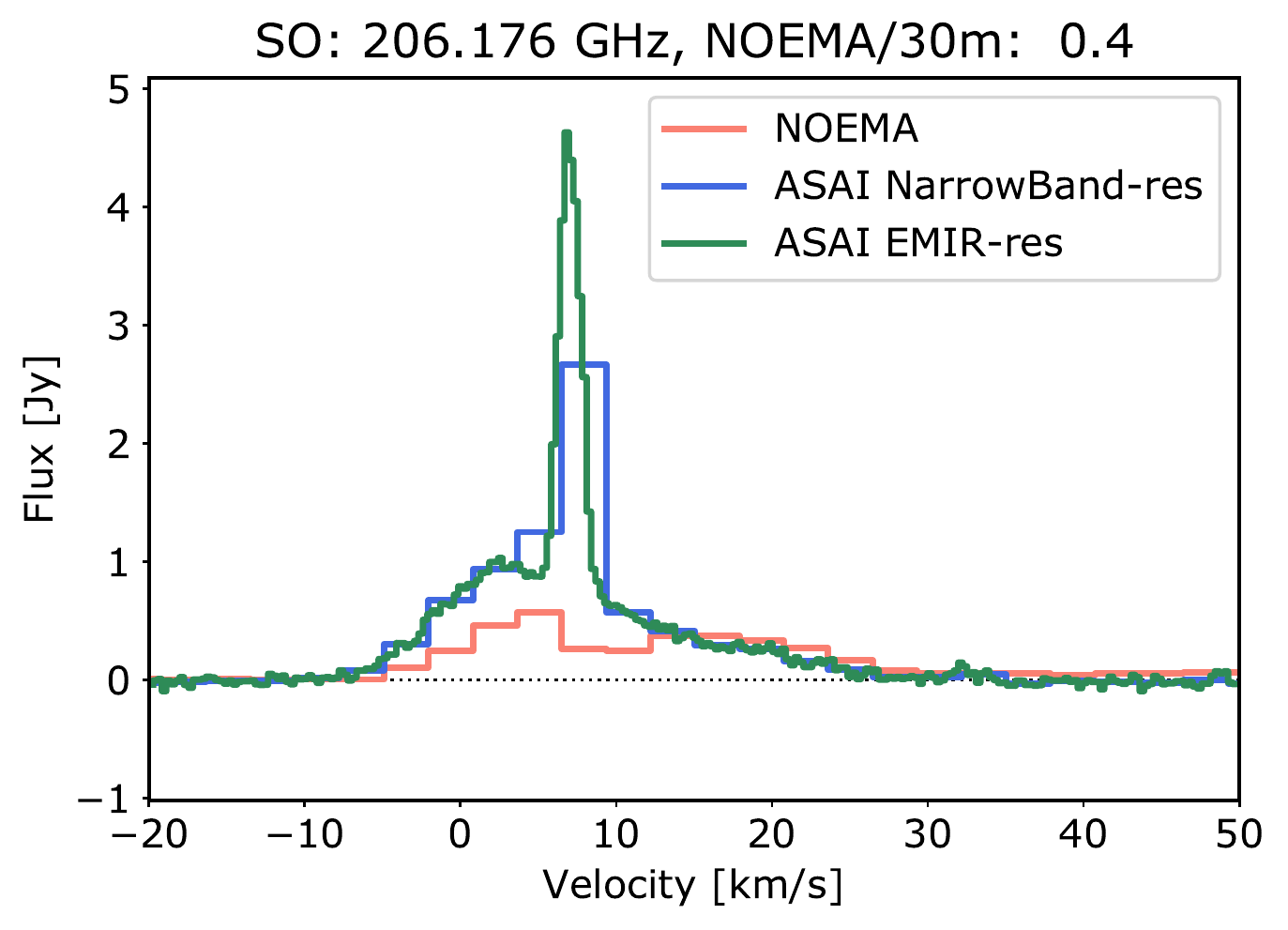}
\includegraphics[width=0.33\textwidth]{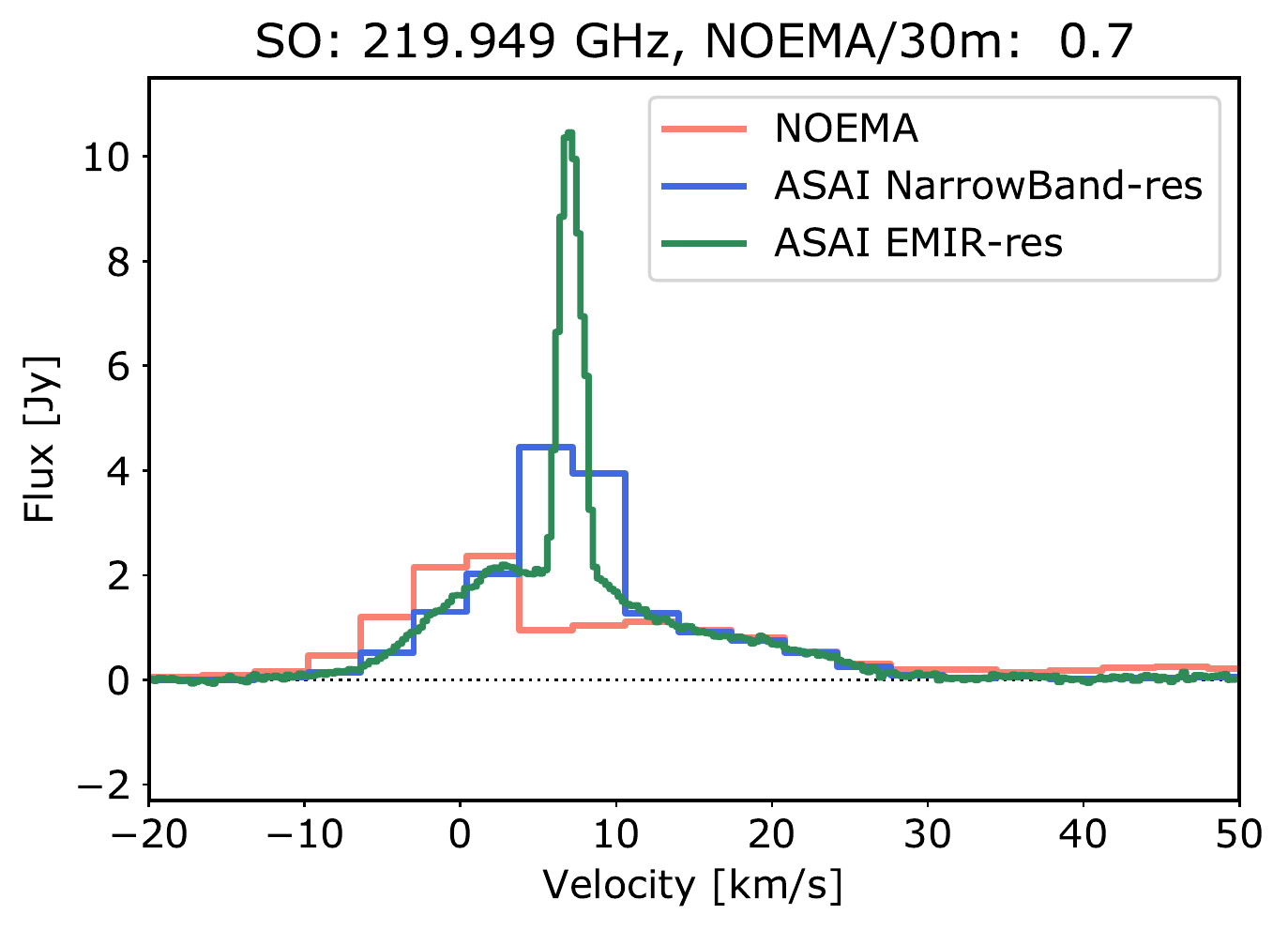} \\
\includegraphics[width=0.33\textwidth]{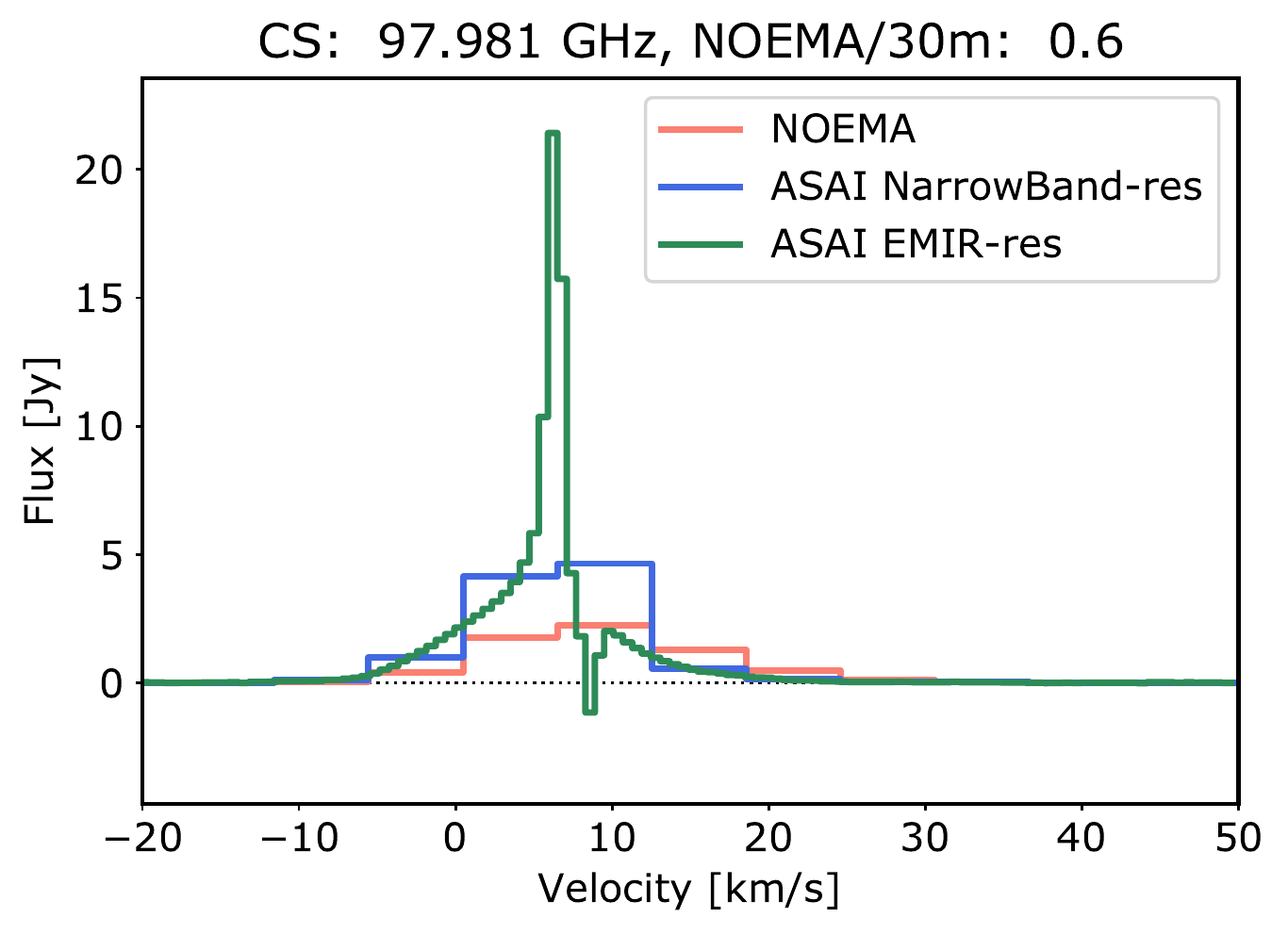} 
\includegraphics[width=0.33\textwidth]{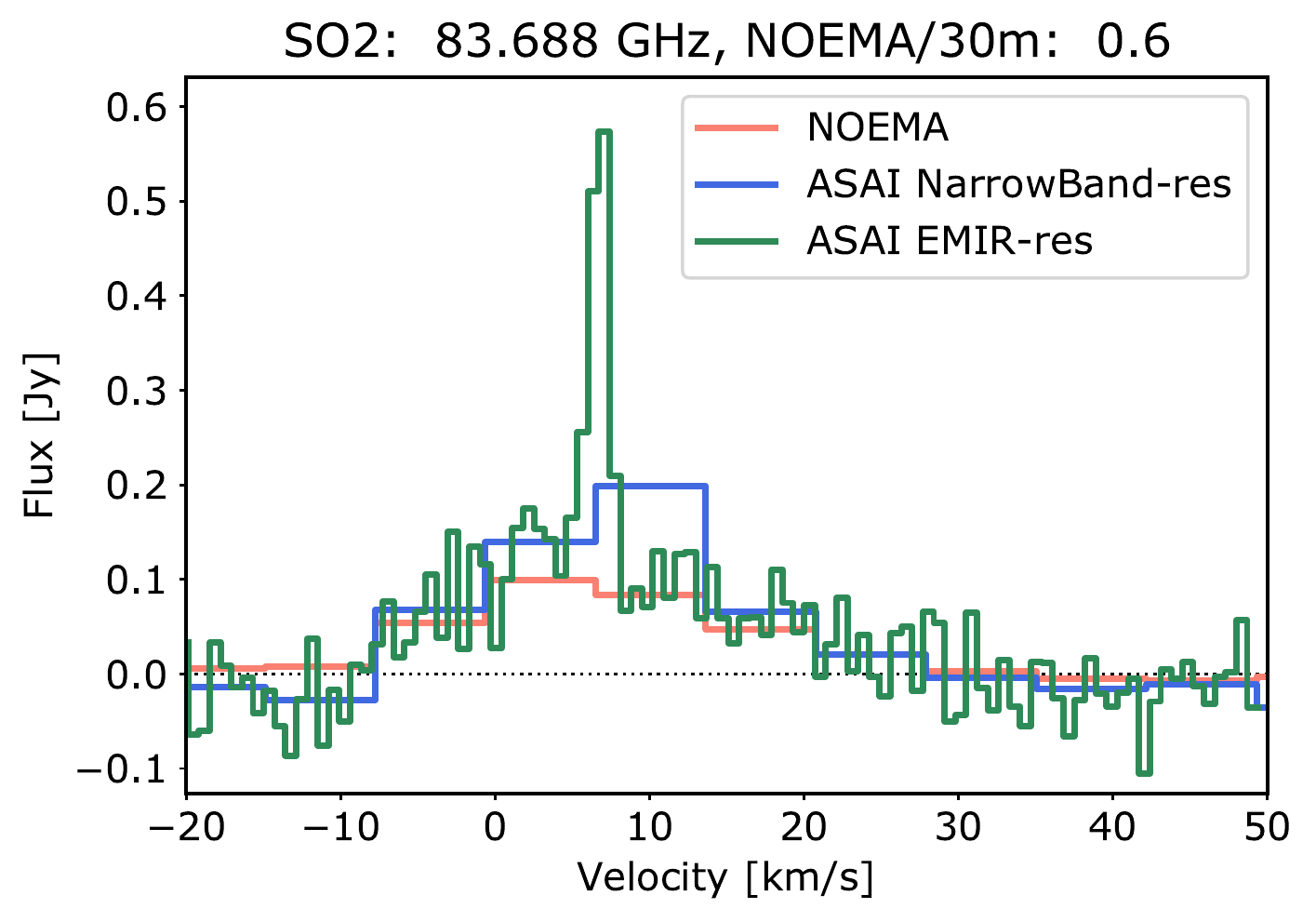}
\includegraphics[width=0.33\textwidth]{spect_mask_flux_SO2_165225.pdf}
\caption{Flux spectra of selected species observed with the IRAM-30m with the ASAI Large Program with the EMIR spectral resolution (in green) and smoothed over the WIDEX resolution (in blue) and with the NOEMA interferometer assuming a mask equal to the IRAM-30m beam (in red). The CH$_3$OH transition has been observed with the NOEMA narrow band backend, the spectrum is here shown at a spectral resolution of 0.4 MHz (0.8 km s$^{-1}$). The flux recovered by NOEMA is reported on top of each panel (see text in section \ref{IRAM30mspectra} for more details). } 
\label{spectra_noema_30m}
\end{figure*}


\begin{thebibliography}{}


\bibitem[Anderson et al.(2013)]{Anderson2013}
{Anderson}, D.~E., {Bergin}, E.~A., {Maret}, S. and {Wakelam}, V. 2013, \apj, 779, 141
\bibitem[Andr{\'e} et al.(2010)]{Andre2010}
Andr{\'e}, P., Men'shchikov, A., Bontemps, S., et al.\ 2010, \aap, 518, L102

\bibitem[Bachiller et al.(1995)]{Bachiller1995}
{Bachiller}, R., {Liechti}, S., {Walmsley}, C.~M. and {Colomer}, F. 1995, \aap, 295, L51
\bibitem[Bachiller et al.(1998)]{Bachiller1998}
{Bachiller}, R., {Codella}, C., {Colomer}, F. et al. 1995, \aap, 335, 266
\bibitem[Bachiller et al.(2001)]{Bachiller2001} 
Bachiller, R., P{\'e}rez Guti{\'e}rrez, M., Kumar, M.~S.~N., et al.\ 2001, \aap, 372, 899
\bibitem[Balan{\c{c}}a et al.(2016)]{Balanca2016} 
Balan{\c{c}}a, C., Spielfiedel, A., \& Feautrier, N.\ 2016, \mnras, 460, 3766
\bibitem[Benedettini et al.(2007)]{Benedettini2007} 
Benedettini, M., Viti, S., Codella, C., et al.\ 2007, \mnras, 381, 1127
\bibitem[Benedettini et al.(2013)]{Benedettini2013} 
Benedettini, M., Viti, S., Codella, C., et al.\ 2013, \mnras, 436, 179
\bibitem[Blitz et al.(2006)]{Blitz2006}
Blitz, M.A., Hughes, K.J., Pilling, M.J., and Robertson, S.H. 2006, J. Phys. Chem. A, 110, 2996
\bibitem[Boogert et al.(1997)]{Boogert1997}
{Boogert}, A.~C.~A., {Schutte}, W.~A., {Helmich}, F.~P., {Tielens}, A.~G.~G.~M., and {Wooden}, D.~H. 1997, \aap, 317, 929
\bibitem[Boogert et al.(2015)]{Boogert2015}
{Boogert}, A.~C.~A., {Gerakines}, P.~A., and {Whittet}, D.~C.~B. 2015, \araa, 53, 541
\bibitem[Bottinelli et al.(2004)]{Bottinelli2004}
 {Bottinelli}, S., {Ceccarelli}, C., {Lefloch}, B. et al. 2004, \apj, 615, 354-358 
 \bibitem[Bottinelli et al.(2007)]{Bottinelli2007}
 {Bottinelli}, S., {Ceccarelli}, C., {Williams}, J.~P., and {Lefloch}, B. 2007, \aap, 463, 601-610
\bibitem[Bron et al.(2018)]{Bron2018} 
Bron, E., Daudon, C., Pety, J., et al.\ 2018, \aap, 610, A12 
\bibitem[Bruderer et al.(2009)]{Bruderer2009} 
Bruderer, S., Benz, A.~O., Doty, S.~D., et al.\ 2009, \apj, 700, 872
\bibitem[Busquet et al.(2014)]{Busquet2014} 
Busquet, G., Lefloch, B., Benedettini, M., et al.\ 2014, \aap, 561, A120
\bibitem[Caselli et al.(1997)]{Caselli1997} 
Caselli, P., Hartquist, T.~W., \& Havnes, O.\ 1997, \aap, 322, 296
\bibitem[Ceccarelli et al.(2017)]{Ceccarelli2017} 
Ceccarelli, C., Caselli, P., Fontani, F., et al.\ 2017, \apj, 850, 176
\bibitem[Charnley(1997)]{Charnley1997}
{Charnley}, S.~B. 1997, \apj, 481, 396
\bibitem[Ching et al.(2016)]{Ching2016}
{Ching}, T.-C., {Lai}, S.-P., {Zhang}, Q., {Yang}, L., {Girart},
J.~M., and {Rao}, R. 2016, \apj, 819, 159
\bibitem[Choi(2005)]{Choi2005}
{Choi}, M. 2005, \apj, 630, 976 
\bibitem[Choi et al.(2007)]{Choi2007}
{Choi}, M., {Tatematsu}, K., {Park}, G. and {Kang}, M. 2007, \apjl, 667, L183
\bibitem[Choi et al.(2011)]{Choi2011}
{Choi}, M., {Kang}, M., {Tatematsu}, K. et al. 2011, \pasj, 63, 1281 
\bibitem[Codella, \& Bachiller(1999)]{Codella1999} 
Codella, C., \& Bachiller, R.\ 1999, \aap, 350, 659
\bibitem[Codella et al.(2005)]{Codella2005} 
Codella, C., Bachiller, R., Benedettini, M., et al.\ 2005, mnras, 361, 244
\bibitem[Codella et al.(2009)]{Codella2009} 
Codella, C., Benedettini, M., Beltr{\'a}n, M.~T., et al.\ 2009, \aap, 507, L25
\bibitem[Codella et al.(2010)]{Codella2010} 
Codella, C., Lefloch, B., Ceccarelli, C., et al.\ 2010, \aap, 518, L112
\bibitem[Codella et al.(2017)]{Codella2017} 
Codella, C., Ceccarelli, C., Caselli, P., et al.\ 2017, \aap, 605, L3
\bibitem[David \& Vassilvitskii(2007)]{David2007}
David, A., \& Vassilvitskii, S. 2007, Proceedings of the eighteenth annual ACM-SIAM symposium on Discrete algorithms, Society for Industrial and Applied Mathematics
\bibitem[De Simone et al.(subm.)]{DeSimone2019}
De Simone, M., Codella, C., Ceccarelli, C. et al.\ 2019, \aap, submitted
\bibitem[Draine \& Salpeter(1979)]{Draine1979} Draine, B.~T., \& Salpeter, E.~E.\ 1979, \apj, 231, 438
\bibitem[Dulieu et al.(2019)]{Dulieu2019} 
Dulieu, F., Nguyen, T., Congiu, E., Baouche, S., and Taquet, V. 2019, \mnras, 484, L119
\bibitem[Flower \& Pineau des Forets(1994)]{Flower1994} Flower, D.~R., \& Pineau des Forets, G.\ 1994, \mnras, 268, 724
\bibitem[Flower \& Pineau des Forêts(2003)]{Flower2003}
{Flower}, D.~R., \& {Pineau Des For{\^e}ts}, G. 2003, \mnras, 343, 390
\bibitem[Flower \& Pineau des For{\^e}ts(2010)]{Flower2010}
{Flower}, D.~R., {Pineau Des For{\^e}ts}, G. and {Rabli}, D. 2010,
\mnras, 409, 29
\bibitem[Flower \& Pineau des For{\^e}ts(2015)]{Flower2015}
{Flower}, D.~R. and {Pineau des For{\^e}ts}, G. 2015, \aap, 578, A63
\bibitem[Geppert et al.(2006)]{Geppert2006} 
Geppert, W.~D., Hamberg, M., Thomas, R.~D., et al.\ 2006, Faraday Discussions, 133, 177
\bibitem[Goldsmith \& Langer(1999)]{Goldsmith1999}
{Goldsmith}, P.~F. and {Langer}, W.~D. 1999, \apj, 517, 209
\bibitem[G{\'o}mez-Ruiz et al.(2015)]{GomezRuiz2015} 
G{\'o}mez-Ruiz, A.~I., Codella, C., Lefloch, B., et al.\ 2015, \mnras, 446, 3346
\bibitem[Green(1995)]{Green1995}
{Green}, S. 1995, \apjs, 100, 213
\bibitem[Gueth et al.(1998)]{Gueth1998} 
Gueth, F., Guilloteau, S., \& Bachiller, R.\ 1998, \aap, 333, 287
\bibitem[Guillet et al.(2007)]{Guillet2007} Guillet, V., Pineau Des For{\^e}ts, G., \& Jones, A.~P.\ 2007, \aap, 476, 263
\bibitem[Gusdorf et al.(2008a)]{Gusdorf2008a} 
Gusdorf, A., Cabrit, S., Flower, D.~R., et al.\ 2008, \aap, 482, 809
\bibitem[Gusdorf et al.(2008b)]{Gusdorf2008b} 
Gusdorf, A., Pineau Des For{\^e}ts, G., Cabrit, S., et al.\ 2008, \aap, 490, 695
\bibitem[Hasegawa \& Herbst(1993)]{hasegawa1993} 
{Hasegawa}, T.~I. and {Herbst}, E. 1993, \mnras, 263, 589-606
\bibitem[Herbst, \& van Dishoeck(2009)]{Herbst2009} 
Herbst, E., \& van Dishoeck, E.~F.\ 2009, \araa, 47, 427
\bibitem[Hirota et al.(2008)]{Hirota2008}
{Hirota}, T., {Bushimata}, T., {Choi}, Y.~K. et al. 2008, \pasj, 60, 37
\bibitem[Holdship et al.(2016)]{Holdship2016} 
Holdship, J., Viti, S., Jimenez-Serra, I., et al.\ 2016, \mnras, 463, 802
\bibitem[Holdship et al.(2018)]{Holdship2019}
Holdship, J., Jimenez-Serra, I., Viti, S., et al.\ 2019, \apj, 878, 64
\bibitem[Kristensen et al.(2012)]{Kristensen2012}
{Kristensen}, L.~E., {van Dishoeck}, E.~F., {Bergin}, E.~A., 
	{Visser}, R., {Y{\i}ld{\i}z}, U.~A., {San Jose-Garcia}, I. et al. 2012, \aap, 2012, 542, A8
\bibitem[Ladjelate et al.(2020)]{Ladjelate2020} 
Ladjelate, B., Andr{\'e}, P., K{\"o}nyves, V., et al.\ 2020, \aap, in press
\bibitem[Laas \& Caselli(2019)]{Laas2019} 
Laas, J.~C., \& Caselli, P.\ 2019, \aap, 624, A108
\bibitem[Leen, \& Graff(1988)]{Leen1988} 
Leen, T.~M., \& Graff, M.~M.\ 1988, \apj, 325, 411
\bibitem[Lef{\`e}vre et al.(2017)]{Lefevre2017} 
Lef{\`e}vre, C., Cabrit, S., Maury, A.~J., et al.\ 2017, \aap, 604, L1
\bibitem[Lefloch et al.(2017)]{Lefloch2017} 
Lefloch, B., Ceccarelli, C., Codella, C., et al.\ 2017, mnras, 469, L73
\bibitem[Lefloch et al.(2018)]{Lefloch2018}
{Lefloch}, B., {Bachiller}, R. et al. 2018 \mnras, 477, 4792
\bibitem[Lloyd(1982)]{Lloyd1982}
Lloyd, S. 1982,  IEEE Transactions on Information Theory, 28, 2
\bibitem[Loison et al.(2012)]{Loison2012}
{Loison}, J.-C., {Halvick}, P., {Bergeat}, A., {Hickson}, K.~M., and {Wakelam}, V. 2012, \mnras, 421, 1476
\bibitem[Looney et al.(2000)]{Looney2000} 
Looney, L.~W., Mundy, L.~G., \& Welch, W.~J.\ 2000, The Astrophysical Journal, 529, 477
\bibitem[L{\'o}pez-Sepulcre et al.(2017)]{LopezSepulcre2017}
{L{\'o}pez-Sepulcre}, A., {Sakai}, N., {Neri}, R. et al. 2017, \aap, 606, A121
\bibitem[Maury et al.(2019)]{Maury2019} Maury, A.~J., Andr{\'e}, P., Testi, L., et al.\ 2019, \aap, 621, A76
\bibitem[May et al.(2000)]{May2000} 
May, P.~W., Pineau des For{\^e}ts, G., Flower, D.~R., et al.\ 2000, \mnras, 318, 809
\bibitem[Nisini et al.(2010)]{Nisini2010} 
Nisini, B., Benedettini, M., Codella, C., et al.\ 2010, \aap, 518, L120
\bibitem[\"{O}berg et al.(2011)]{Oberg2011} 
\"{O}berg, K.~I., Boogert, A.~C.~A., Pontopiddan, K.~M., et al. 2011, \apj, 740, 109 
\bibitem[Ospina-Zamudio et al.(2018)]{Ospina-Zamudio2018} 
Ospina-Zamudio, J., Lefloch, B., Ceccarelli, C., et al.\ 2018, \aap, 618, A145
\bibitem[Palumbo et al.(1997)]{Palumbo1997}
Palumbo, M. E., Geballe, T. R., Tielens, A. G. G. M. 1997, \apj, 479, 839
\bibitem[Persson et al.(2012)]{Persson2012} 
Persson, M.~V., J{\o}rgensen, J.~K., \& van Dishoeck, E.~F.\ 2012, \aap, 541, A39
\bibitem[Persson et al.(2014)]{Persson2014} 
Persson, M.~V., J{\o}rgensen, J.~K., van Dishoeck, E.~F., et al.\ 2014, \aap, 563, A74
\bibitem[Pineau des Forets et al.(1993)]{PineaudesForets1993} 
Pineau des Forets, G., Roueff, E., Schilke, P., et al.\ 1993, \mnras, 262, 915
\bibitem[Podio et al.(2014)]{Podio2014} 
Podio, L., Lefloch, B., Ceccarelli, C., et al.\ 2014, \aap, 565, A64
\bibitem[Pontoppidan et al.(2004)]{Pontoppidan2004}
{Pontoppidan}, K.~M., {van Dishoeck}, E.~F. and {Dartois}, E. 2004, \aap, 426, 925-940
\bibitem[Rabli \& Flower(2010)]{Rabli2010}
{Rabli}, D. and {Flower}, D.~R. 2010, \mnras, 406, 95 
\bibitem[Santangelo et al.(2014)]{Santangelo2014}
{Santangelo}, G., {Nisini}, B., {Codella}, C., {Lorenzani}, A., 
	{Y{\i}ld{\i}z}, U.~A., {Antoniucci}, S., {Bjerkeli}, P., 
	{Cabrit}, S., {Giannini}, T., {Kristensen}, L.~E., 
	{Liseau}, R., {Mottram}, J.~C., {Tafalla}, M., and {van
          Dishoeck}, E.~F. 2014, \aap, 568, A125
\bibitem[Santangelo et al.(2015)]{Santangelo2015}
{Santangelo}, G., {Codella}, C., {Cabrit}, S. et al. 2015, \aap, 584, A126
\bibitem[Schofield(1973)]{Schofield1973}
Schofield, K. 1972, J. Phys. Chem. Ref. Data, 2, 25
\bibitem[Smith et al.(1991)]{Smith1991}
{Smith}, R.~G. 1991, \mnras, 249, 172
\bibitem[Smith et al.(2004)]{Smith2004}
{Smith}, I.~W.~M., {Herbst}, E., and {Chang}, Q. 2004, \mnras, 350, 323
\bibitem[Tabone et al.(2017)]{Tabone2017} 
Tabone, B., Cabrit, S., Bianchi, E., et al.\ 2017, \aap, 607, L6
\bibitem[Taquet et al.(2012)]{Taquet2012} 
Taquet, V., Ceccarelli, C., \& Kahane, C. 2012, \aap, 538, A42
\bibitem[Taquet et al.(2013)]{Taquet2013}
Taquet, V., Peters, P. , Kahane, C., Ceccarelli, C., L\'opez-Sepulcre, A., Toubin, C., Duflot, D., Faure, A. and Wiesenfeld, L. 2013, \aap, 550, A127
\bibitem[Taquet et al.(2014)]{Taquet2014} 
Taquet, V., Charnley, S. B., Sipil{\"a}, O. 2014, \apj, 791, 1 
\bibitem[Taquet et al.(2015)]{Taquet2015}
Taquet, V., L{\'o}pez-Sepulcre, A., Ceccarelli, C. et al. 2015, \apj, 804, 81 
\bibitem[Taquet et al.(2016)]{Taquet2016}
Taquet, V., Wirstr{\"o}m, E., Charnley, S. B. 2016, \apj, 821, 46
\bibitem[Tielens et al.(1991)]{Tielens1991}
{Tielens}, A.~G.~G.~M., {Tokunaga}, A.~T., {Geballe}, T.~R. and 
	{Baas}, F. 1991, \apj, 381, 181-199
\bibitem[Tielens et al.(1994)]{Tielens1994} Tielens, A.~G.~G.~M., McKee, C.~F., Seab, C.~G., et al.\ 1994, \apj, 431, 321
\bibitem[Tobin et al.(2018)]{Tobin2018} 
Tobin, J.~J., Looney, L.~W., Li, Z.-Y., et al.\ 2018, \apj, 867, 43
\bibitem[Tsunashima et al.(1975)]{Tsunashima1975}
Tsunashima, S., Yokota, T., Safarik, I., Gunning, H.E., Strausz, O.P. 1975, J. Phys. Chem. 1975, 79, 8, 775-778
\bibitem[van der Tak et al.(2007)]{vanderTak2007}
Van der Tak, F.F.S., Black, J.H., Schöier, F.L., Jansen, D.J., and van
Dishoeck, E.F. 2007, \aap 468, 627 
\bibitem[Vidal et al.(2018)]{Vidal2018}
{Vidal}, T.~H.~G., \& {Wakelam}, V. 2018, \mnras, 474, 5575
\bibitem[Viti et al.(2011)]{Viti2011}
{Viti}, S., {Jimenez-Serra}, I., {Yates}, J.~A. et al. 2011, \apjl,
740, L3
\bibitem[Visser et al.(2009)]{Visser2009} 
Visser, R., van Dishoeck, E.~F., Doty, S.~D., et al.\ 2009, \aap, 495, 881
\bibitem[Wakelam et al.(2004)]{Wakelam2004}
{Wakelam}, V., {Caselli}, P., {Ceccarelli}, C., {Herbst}, E., and {Castets}, A. 2004, \aap, 422, 159
  \bibitem[Wakelam, \& Herbst(2008)]{Wakelam2008} 
  Wakelam, V., \& Herbst, E.\ 2008, \apj, 680, 371
\bibitem[Watanabe, \& Kouchi(2002)]{Watanabe2002} 
Watanabe, N., \& Kouchi, A.\ 2002, \apjl, 571, L173
\bibitem[Wilson \& Rood(1994)]{Wilson1994}
Wilson, T. L. and Rood, R. T. 1994, \araa, 32, 191
\bibitem[{Y{\i}ld{\i}z} et al.(2012)]{Yildiz2012}
{Y{\i}ld{\i}z}, U.~A., {Kristensen}, L.~E., {van Dishoeck}, E.~F.,
	{Belloche}, A., {van Kempen}, T.~A., {Hogerheijde}, M.~R.,
	{G{\"u}sten}, R. and {van der Marel}, N. 2012, \aap, 542, A86
\bibitem[Zucker et al.(2018)]{Zucker2018} 
Zucker, C., Schlafly, E.~F., Speagle, J.~S., et al.\ 2018, \apj, 869, 83

 \end{thebibliography}
\end{document}